\documentclass[9pt,twocolumn,twoside]{pnas-new}

\templatetype{pnasresearcharticle} 

\usepackage{amssymb,amsfonts,amsmath}
\usepackage{amssymb,amsfonts,amsmath}
\usepackage[outdir=./]{epstopdf}

\setboolean{displaywatermark}{false}

\usepackage{mathtools}

\renewcommand\Re{\operatorname{Re}}
\renewcommand\Im{\operatorname{Im}}

\DeclareMathOperator{\Prob}{Pr}

\renewcommand*{\vec}[1]{{#1}}

\title{Counting equilibria of large complex systems by instability index}

\author[a,1]{G\'{e}rard Ben Arous}
\author[b,c,1]{Yan V Fyodorov}
\author[d,1]{Boris A Khoruzhenko}

\affil[a]{Courant Institute of Mathematical Sciences, New York University, 251 Mercer Street, New York, NY10012, USA}
\affil[b]{Department of Mathematics, King's College London, London  WC2R 2LS, United Kingdom}
\affil[c]{L. D. Landau Institute for Theoretical Physics, Semenova 1a, 142432 Chernogolovka, Russia}
\affil[d]{Queen Mary University of London, School of Mathematical Sciences, London E1 4NS, United Kingdom}

\significancestatement{This paper sheds light on the phase  portrait structure
of a 'minimal model' for a large nonlinear system of many randomly interacting degrees of freedom equipped with a stability feedback mechanism. In this way it significantly
 extends local stability analysis of large complex systems which was undertaken by Robert May in 1972.  We show that the transition from stability to instability is characterized by the exponential explosion in the number of unstable equilibria, with drastically reduced probability of finding truly locally attracting equilibria . At the same time we demonstrate abundance of equilibria with a large proportion of stable directions, which arguably can slow down the system dynamics for sufficiently long time and induce aging effects.}

\authorcontributions{(1) Author contributions: G.B.A., Y.V.F. and B.A.K. performed research and wrote the paper}
\authordeclaration{The authors declare no conflict of interest.}
\correspondingauthor{\vspace*{-2ex}Corresponding Authors:\\ benarous@cims.nyu.edu, yan.fyodorov@kcl.ac.uk, b.khoruzhenko@qmul.ac.uk}

\keywords{complex systems $|$ stability $|$ equilibrium $|$ random matrices }

\begin{abstract}
{We consider a nonlinear autonomous system of $N\gg 1$ degrees of freedom
randomly coupled by both relaxational ('gradient') and non-relaxational ('solenoidal') random  interactions. We show that with increased interaction strength such systems generically undergo an abrupt transition
from a trivial phase portrait with a single stable equilibrium into a topologically non-trivial regime of 'absolute instability'
where equilibria are on average exponentially abundant, but typically all of them are unstable, unless the dynamics is purely gradient. When interactions increase  even further the stable equilibria eventually become on average exponentially abundant unless the interaction is purely solenoidal.  We further calculate the mean proportion of equilibria which have a fixed fraction of unstable directions.
}
\end{abstract}

\dates{This manuscript was compiled on \today}
\doi{\url{www.pnas.org/cgi/doi/10.1073/pnas.XXXXXXXXXX}}

\begin{document}

\maketitle
\thispagestyle{firststyle}
\ifthenelse{\boolean{shortarticle}}{\ifthenelse{\boolean{singlecolumn}}{\abscontentformatted}{\abscontent}}{}

\dropcap{I}n 1972, in his seminal paper \cite{May1972}  Robert May analyzed the relationship between complexity and stability of large complex systems at equilibrium. Although May was motivated by the "stability versus diversity" debate in ecology \cite{AlessinaRev}, his neighborhood stability analysis
applies far beyond model ecology, e.g., to neural networks
\cite{SCS1988,WT}, systemic risk in trading
 \cite{FS13} or modeling of large economies \cite{MB2019}. To recap it, 
 consider a system of $N\gg 1 $ degrees of freedom $\vec{x}=(x_1,\ldots, x_N)^T$, whose evolution is governed by a set of coupled non-linear first-order ordinary differential equations.
The local stability analysis of an equilibrium, say $\vec{x}_*$,  amounts to linearizing the system near $\vec{x}_*$
and looking at the time evolution of
the displacement 
$\vec{y}=\vec{x}-\vec{x}_*$.
Assuming that each of the degrees of freedom $x_i$ by itself, when disturbed from equilibrium, returns back with some characteristic time independent of $i$, such evolution is described by the equation $\dot{\vec{y}}=-\mu\vec{y} + J\vec{y} $. 
Here, the parameter $\mu>0$ sets the characteristic relaxation time in the absence of interactions and the matrix $J=(J_{ij})$ describes the pair-wise interactions between the degrees of freedom in the neighborhood of $\vec{x}_*$. 

To get insights into the interplay between stability and complexity,
May simplified the problem by assuming that positive and negative values of the pair-wise interactions $J_{ij}$ are equally likely to occur,
a plausible assumption for large complex systems. Accordingly, he chose $J_{ij}$  to be random variables with zero mean and 
standard deviation $\sigma$ (`typical' interaction strength), thus retaining the fewest possible number of control parameters in his model.
Invoking random matrix theory,
he then concluded that large complex systems exhibit a sharp transition from stability to instability when the number of degrees of freedom or the interaction strength increase beyond
the instability threshold which is given by a remarkably simple equation $\mu=\sigma\sqrt{N}$. 

One obvious limitation of the neighborhood stability analysis is that it gives no insight
into what happens outside the immediate neighborhood of equilibrium when it becomes unstable.
Hence, May's analysis has only limited bearing on the dynamics of populations operating out-of-equilibrium \cite{AlessinaRev}.
For example, in the context of model ecosystems, populations may coexist thanks to limit cycles or chaotic attractors, which typically originate from unstable equilibrium points. This naturally prompts important lines of enquiry for large complex systems about classification of equilibria by stability, studying basins of attraction, and other features of global dynamics. 

In an extension of May's work, two of us introduced a `minimal' nonlinear model of large complex
systems equipped with a stability feedback mechanism 
\cite{FyoKhor2016}. The main finding of \cite{FyoKhor2016} was that such systems
 exhibit a
 transition from a trivial phase portrait with a single stable equilibrium to one characterized by exponentially many equilibria.
 However, the important  question
about stability of those exponentially many equilibria remained unanswered. 
In the present paper we develop a framework 
for a statistical description of equilibria of large complex systems and then use it to calculate frequencies of stable equilibria and, also, of equilibria with a fixed fraction of stable directions. 

Statistics of unstable equilibria with a large fraction of stable directions are of a particular interest in the context of large complex systems with an underlying energy landscape. In that case the dynamics can be visualized as a gradient descent on the energy surface, and, as was argued in ref.~\cite{Kurchan_1996} the system is trapped near borders (ridges) of basins of attraction of local minima because of the dominance of borders in large-dimensional spaces. The gradient descent is then determined mainly by nearby saddles  which lie on the ridges, which may trap dynamics for a long time due to the large number of stable directions. It is natural to expect that unstable equilibria with a large number of attracting directions may play a similar role in non-gradient dynamics, providing a motivation for our research.

\subsection*{Statistics of equilibria}
The model studied in \cite{FyoKhor2016} is described by a system of autonomous non-linear differential equations
 \begin{equation}\label{1}
 \dot{\vec{x}}= -\mu \vec{x} + \vec{f} (\vec{x}), \quad \vec{x}\in \mathbb{R}^N,
 \end{equation}
 coupled via a smooth random vector field $\vec{f}(\vec{x})$
which models  both the complexity and nonlinearity of interactions.
Finding equilibria, i.e. solutions of Eq.~\ref{1} which do not change with time, amounts to solving
the equation $-\mu \vec{x} +\vec{f} (\mathbf{x})=0$.
Since the interaction field $\vec{f}(\vec{x})$ is random,  the total number of equilibria and their locations are not fixed in our model and
may change from one realization of $\vec{f}(\vec{x})$ to another.
Thus, in contrast to the neighborhood stability analysis of a known equilibrium which was carried out by May, our model does not provide insights into properties of a single given equilibrium. Instead it makes possible a statistical analysis of stability properties of equilibria.
Effectively,  May's question ``Will a large complex system be stable'' in our model is replaced by the question "What is the probability that an equilibrium drawn at random from the entire population of equilibria is stable?"

This probability, denote it  by $p_{st}$, can be written in terms of counting statistics of equilibria.
If ${\cal N}_{eq}$ is the total number of equilibria and ${\cal N}_{st}$ is the number of stable equilibria, then
one can argue, see below, that
\begin{equation}\label{p_st1}
p_{st}=\langle {\cal N}_{st}/{\cal N}_{eq}\rangle \, ,
\end{equation}
where the angle brackets stand for averaging over $\vec{f}(\vec{x})$.

Both counting functions, ${\cal N}_{eq}$ and ${\cal N}_{st}$, are examples of linear statistics of equilibria
of the form
\begin{equation*}\label{LS}
\mbox{$L[\mathit{\Psi}]=\sum_{\vec{x}_*}\!\! \mathit{\Psi}(\vec{x}_*)$}
\end{equation*}
where the sum is over all equilibria and $\Psi$ is a test function. In these
notations,
${\cal N}_{eq}=L[1]$ and  ${\cal N}_{st}=L\big[\mathit{\Theta} (\mu - x_{max}(J))\big]$,
where  $J$ is the Jacobian matrix $({\partial f_i}/{\partial x_j})$ of the vector field $\vec{f}(\vec{x})$, $x_{max}(J)$ is the largest real part of the eigenvalues of $J$
and
$\mathit{\Theta} (x)$ is the Heaviside step function, so that $\mathit{\Theta} (\mu - x_{max}(J))$ is the indicator-function of the event that $x_{max}(J)<\mu$.

The test function
$\mathit{\Psi}_J\! =\!\delta (J \!- \!({\partial f_i}/{\partial x_j} ))$, where
$\delta (J)$ is the matrix delta-function provides another 
example of linear statistics of equilibria.
Its weighted average,
\[
P_{eq} (J) =
\big\langle
     \frac{1}{{\cal N}_{eq}} \sum_{\vec{x}_*}
     \prod_{i,j}  \delta
           \big(
                J_{ij} - \frac{\partial f_i}{\partial x_j} (\vec{x}_*)
           \big)
 \big\rangle \, ,
\]
is {\it the sample mean}  of the joint probability density function for the matrix elements $J_{ij}$ of the Jacobian at equilibrium.
Then, the probability for a randomly selected equilibrium point to be stable is
$
p_{st}\!= \!\int\!\!  \mathit{\Theta} (\mu \! -\! x_{max} (J)) P_{eq} (J)\, dJ \, .
$
On replacing $P_{eq} (J)$ here with its expression in terms of the weighted average above, one immediately obtains Eq.~\ref{p_st1}.

One can extend this statistical framework
from the binary descriptor of points of equilibria (stable or unstable) to a continuous one. Define $\kappa (\vec{x}_*)$ to be the dimension of the local unstable manifold  of the non-linear system [\ref{1}] at equilibrium $\vec{x}_*$, i.e., $\kappa (\vec{x}_*)$ is the number of eigenvalues of the  matrix $-\mu \delta_{ij} + \frac{\partial f_i}{\partial x_j}$ at $\vec{x} =\vec{x}_*$ with positive real parts. In the limit $N\gg 1$, the fraction $\kappa (\vec{x}_*)/N$ 
 can be interpreted as a measure of instability of the equilibrium at $\vec{x}_*$. We shall call an equilibrium $\alpha$-stable if its {\it instability index} $\kappa/N$ does not exceed value $\alpha$
and denote by ${\cal N}_{st}^{(\alpha)}$ the number of $\alpha$-stable equilibria. Then the probability
 that an equilibrium drawn at random from the entire population of equilibria will
have its instability index in the interval $(\alpha_1, \alpha_2)$ is $\int_{\alpha_1}^{\alpha_2} \nu({\alpha}) \, d\alpha$, where
 \begin{equation}\label{quenched}
\nu({\alpha})={dp_{\alpha}}/{d\alpha}, \quad p_{\alpha} =
\langle
{\cal N}_{st}^{(\alpha)}\!/{\mathcal{N}_{eq}}
 \rangle\, .
\end{equation}
The counting function ${\cal N}_{st}^{(\alpha)}$ can too be cast in the framework of linear statistics of equilibria. To this end, let us order the eigenvalues $z_j$ of the Jacobian matrix $J=({\partial f_i}/{\partial x_j})$ by their real parts $x_j=\Re z_j$ so that $x_{max}=x_1\ge x_2 \ge \ldots \ge x_N$\footnote{The matrix $J$ is random and, typically eigenvalues of such matrices are all distinct. Therefore, this labelling (ordering arrangement) is consistent}. Then ${\cal N}_{st}^{(\alpha)}=L[\mathit{\Theta} (\mu-x_{\alpha N+1})]$. 

The computation of $p_{st}$ and $\nu (\alpha)$
is a challenging problem. Instead, in this paper we study their `annealed' versions,
 \begin{equation}\label{denind}
p^{(a)}_{st}=\frac{\langle {\cal N}_{st}\rangle}{\langle {\cal N}_{eq}\rangle}, \quad \nu^{(a)}_N(\alpha)=\frac{1}{\langle \mathcal{N}_{eq}\rangle}
\frac{d}{d\alpha}\langle \mathcal{N}^{(\alpha)}_{st}\rangle\, ,
\end{equation}
thus reducing the problem to calculating the expected number of stable and $\alpha$-stable equilibria. 
These are interesting observables on their own.  As we shall shaw below the corresponding annealed complexity exponents, see Eq. \ref{S_alpha}, which can be computed in a closed form in the limit of large number of degrees of freedom, exhibit not-trivial dependences on the model parameters. Also, in the particular case of gradient flow $ \dot{\vec{x}}= -\nabla L(\vec{x})$, the annealed complexity exponents of local minima and other stationary points on the  (random) surface of the potential function $L(\vec{x})$ attracted considerable recent interest 
and especially in the context of glassy dynamics. 
Still, the question of whether the annealed probabilities [\ref{denind}] give any insight into their quenched counterparts [\ref{quenched}] is an important open question. 
The recent progress \cite{Subag2017,RBC2019,Auf2020b} in understanding this question in the context of gradient dynamics on the energy surface of the $p$-spin spherical model gives rise to a hope that for some classes of coupling fields in [\ref{1}] the annealed picture will resemble the quenched one. The task of identifying such classes of coupling fields is a challenging open problem which deserves further investigation along with the companion question about the qualitative differences between the quenched and annealed pictures.

\subsection*{Model assumptions} To get insights into statistics of equilibria of large complex systems, we  follow the philosophy of the '`minimal' model \cite{FyoKhor2016} and decompose   the coupling field  into the sum of gradient (curl-free) and solenoidal (divergence-free) components:
\begin{equation}\label{2}
  f_i(\vec{x})=-\frac{\partial V(\vec{x})}{\partial x_i}+\frac{1}{\sqrt{N}}\sum_{j=1}^N\frac{\partial A_{ij}(\vec{x})}{\partial x_j}, \; i=1,\ldots,N,
 \end{equation}
where the matrix $A(\vec{x})$
 is antisymmetric: $A_{ij}(\vec{x})=-A_{ji}(\vec{x})$
for every $\vec{x}$.
Such a representation  provides a rich, though not most general, class of vector fields
 The scalar and vector potentials,  $V(\vec{x})$ and $A(\vec{x})$ respectively, are assumed to be statistically independent, zero mean Gaussian random fields, with smooth realizations and the additional assumptions of  {\it homogeneity} (translational invariance) and {\it isotropy} (rotational invariance):
\begin{eqnarray}\label{3a}
\langle V(\vec{x}) V(\vec{y})\rangle&\!\!\!=\!\!\!&v^2 \mathit{\Gamma}_V\! \left(|\vec{x}-\vec{y}|^2 \right)\, , \\
\label{3b}
\hspace*{-3ex} \langle A_{ij}(\vec{x}) A_{nm}(\vec{y})\rangle&\!\!\!=\!\!\!&a^2 \mathit{\Gamma}_A\! \left(|\vec{x}-\vec{y}|^2\right) \! (\delta_{in}\delta_{jm}\!-\!\delta_{im}\delta_{jn})\, .
\end{eqnarray}
The covariance functions, $ \mathit{\Gamma}_V$ and $ \mathit{\Gamma}_A$ are normalised by the condition $\frac{d^2}{ds^2}\mathit{\Gamma}_{V,A}(s)\left|_{s=0} \right. =1$. 
The covariance functions of isotropic Gaussian fields were first studied in \cite{S1938}. In particular, for $\mathit{\Gamma} \left(|\vec{x}-\vec{y}|^2 \right)$ to define a covariance function in all dimensions $\mathbb{R}^N$ it must hold that 
$\mathit{\Gamma} (t)=\int_0^{\infty} \exp ({- s t}) dG(s)$, $\forall t\ge 0$, 
for some finite measure $dG(s)$ on $\mathbb{R}_{+}$.

Our model has the fewest possible number of parameters. These are
\begin{equation}\label{deftau}
\tau=\frac{v^2}{v^2+a^2} \quad \mbox{and}\quad m=\frac{\mu}{\sqrt{4N(v^2+a^2)}}. 
\end{equation}
The scaled relaxation strength $m$ is a measure of the strength of the stability feedback mechanism relative to the interaction strength and the \emph{potentiality parameter} $\tau$ controls the balance between the gradient and solenoidal components of the interaction. If $\tau=1$ then the flow defined by Eq.~\ref{1} is purely gradient: $  \dot{\vec{x}}= -\nabla L (\vec{x}) $, with $L (\vec{x}) = \mu |\vec{x}|^2/2 - V(\vec{x})$ being the associated Lyapunov function. And if $\tau=0$ then the interaction field $\vec{f}(\vec{x})$ is divergence free. 

Note that 
$m$ is essentially the same control parameter as one in May's linear model.  In the non-linear setting, it controls the complexity of the phase portrait.  As was shown in ref.~\cite{FyoKhor2016}, for large values of $m$ the stability feedback mechanism prevails and, typically,  the system has a single equilibrium which is stable. When the value of $m$ decreases, the system exhibits a sharp transition from this simple phase portrait to a complex one which is characterized by exponentially growing number of equilibria. More precisely, to leading order in the limit $N\gg 1 $, 
\begin{eqnarray}\label{N_eq}
\langle {\cal N}_{eq}\rangle =
\begin{cases}
1, & \mbox{if } m>1 , \\[1ex]
\sqrt{\frac{2(1+\tau)}{1-\tau}} \, e^{N\mathit{ \Sigma}_{eq} (m)} \, , &  \mbox{if } 0< m<1 \, ,
\end{cases}
\end{eqnarray}
where
\begin{equation*}\label{Sigma_eq}
\mathit{\Sigma}_{eq}(m) = \frac{1}{2} (m^2-1) -\ln m\, .
\end{equation*}
Thus, as far as the total number of equilibria
is concerned, the picture that is emerging in the limit  $N\gg 1$ is largely independent of $\tau$, although the pre-exponential factor in Eq.~\ref{N_eq} suggests that the case of pure gradient flow $\tau=1$ is special
\footnote{
In this case the task of counting (and classifying) equilibria is equivalent to counting saddle-points, minima and maxima of random potentials, see discussion in \cite{FyoKhor2016}. That counting has been done earlier by several methods \cite{Fyo04,BD07,FyoWi07,FyoNad2012,Auf1,Auf2}, see also \cite{BNM2017,Ros2018,R2020}. Within the confines of model [\ref{1}], the pure gradient flow can be approached in the weakly non-gradient limit  $\tau=1-u^2/N$ \cite{FyoKhor2016}.}.

The role of potentiality parameter $\tau$ will be revealed by our subsequent analysis.  

Coming back to our model assumptions, one can use a different class of coupling fields $f(x)$ - homogenous Gaussian fields with zero mean and covariance  
\[
\langle f_i(\vec{x} )f_j (\vec{y} )  \rangle= \mathit{\Gamma}_1(|x-y|^2)\delta_{ij} + \mathit{\Gamma}_2(|x-y|^2)(x_i-y_i)(x_j-y_j)\, .
\]
This class of  random fields have been used in the statistical theory of isotropic turbulence since 1930s \cite{KH1938,R1940}
and in the context of large complex systems very recently in ref. \cite{Ipsen2017}.  Although such fields have a different covariance structure to the one given by Eqs \ref{3a}--\ref{3b}, our stability analysis extends to this class  almost verbatim. 

One can also consider inhomogenous coupling fields, see ref.~\cite{FFI2021}. 
 As far as the assumptions of isotropy and Gaussianity of the coupling field $\vec{f} (\vec{x})$ are concerned, recent progress in the evaluation of the rate of growth of random determinants of large random matrices with non-invariant matrix distribution  \cite{BBMcK1, BBMcK2} raise hope that the isotropy assumption could be relaxed, at least in the gradient case. However, the Gaussianity assumption is indispensable. This assumption allows one to compute the Kac-Rice integral via its reducing to conditional averages of random matrix determinants, see Materials and Methods, and  without it an effective computation of the Kac-Rice integral, Eq. [\ref{KC_int}], seems hardly possible. 
 
\subsection*{Stable~equilibria~and~stable~directions~of~unstable~equilibria}The
starting point of our analysis of stability properties of equilibria is the
Kac-Rice formula for counting solutions of simultaneous equations. 
By expressing the mean value of linear statistics of equilibria $\langle L[\mathit{\Psi}]\rangle $ as a random matrix average, it brings the original counting problem into the realms of random matrix theory,
see ref.~\cite{FyoKhor2016}.
If $\mathit{\Psi} (\vec{x}_*)=\psi (J_*)$, where $J_*=({\partial f_i}/{\partial x_j})$ is the Jacobian matrix of the interaction field $\vec{f}(\vec{x}) $ at $\vec{x}=\vec{x}_*$, 
then (see Section Materials and Methods)
\begin{equation}\label{LS1}
\langle L[\mathit{\Psi}]  \rangle_{\vec{f}} = \mu^{-N} \langle \psi (J) \, |\!\det(-\mu I +J )|\rangle_J\, .
\end{equation}
Here the angle brackets on the left-hand side stand for the averaging over realizations of the interaction field $\vec{f}(\vec{x})$, and the angle brackets on the right-hand side stand for the averaging over the distribution of the Jacobian matrix $J$. The latter does not depend of $\vec{x}$ because of the homogeneity of $\vec{f}(\vec{x})$.

\begin{figure}[t!]
\centering
\includegraphics[width=.8\linewidth]{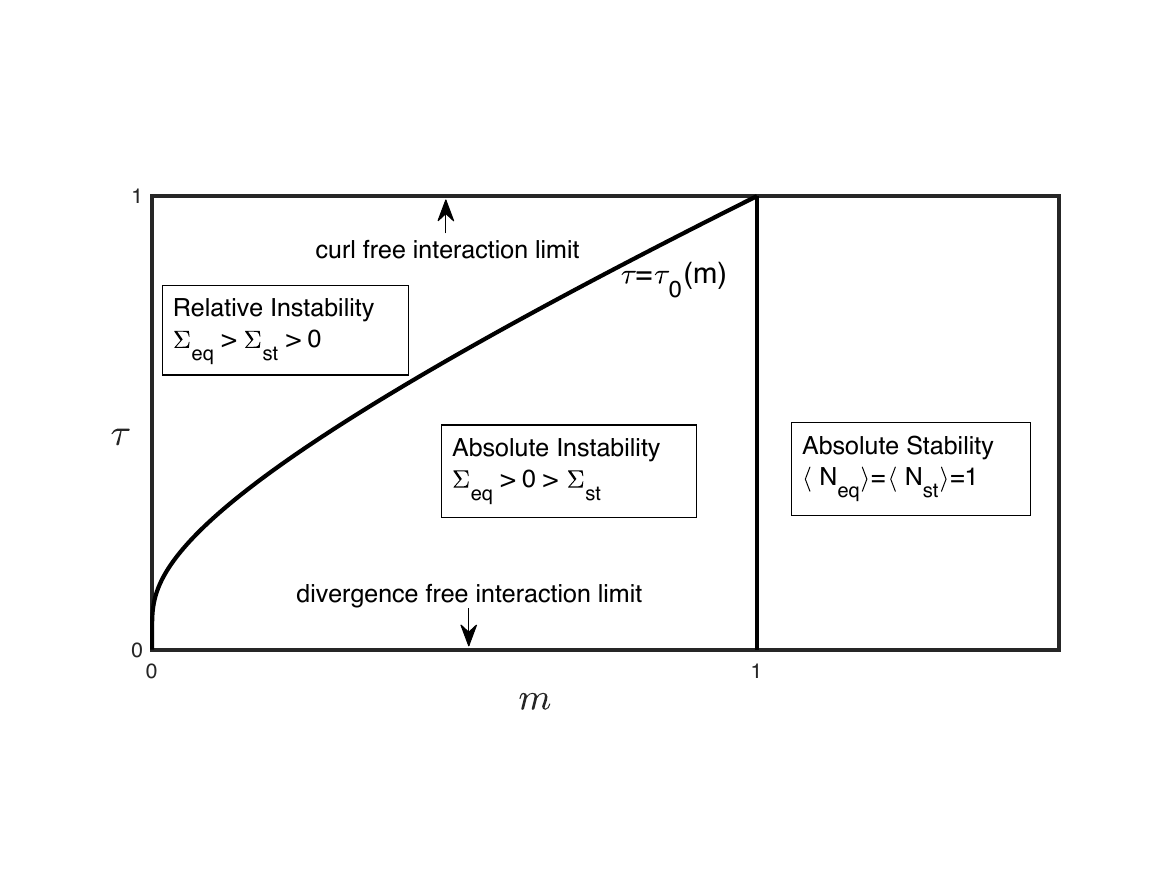}
\vspace*{-6ex}
\caption{
\label{fig}
 Phase diagram of model [\ref{1}]. The curve $\tau=\tau_0(m)$ separates the regions of absolute and relative instability in the `topologically non-trivial' phase.
}
\end{figure}

Eq.~\ref{LS1} makes it possible to draw on analytic techniques from random matrix theory and compute counting statistics of equilibria, such as the complexity exponent $\mathit{\Sigma}_{eq} (m) $ associated with the total number of equilibria. 
 In this context, powerful tools of  {\it Large Deviation Theory} developed for matrices with complex eigenvalues in refs.~\cite{BG1997,BZ1998} become especially useful.
They allow one to compute the complexity exponents $\mathit{\Sigma}_{st}$ and $ \mathit{\Sigma}^{(\alpha)}_{st}$ associated with the stable and $\alpha$-stable equilibria:
\begin{equation}\label{S_alpha}
\mathit{\Sigma}_{st}= \lim_{N\to\infty}  \frac{1}{N} \ln \langle {\cal N}_{st}\rangle, \quad  \mathit{\Sigma}^{(\alpha)}_{st} = \lim_{N\to\infty}  \frac{1}{N} \ln \langle {\cal N}_{st}^{(\alpha)}\rangle \, .
\end{equation}
One outcome of this computation (see Supplemental Information)
is a closed form expression for $\mathit{\Sigma}_{st}$ in the topologically nontrivial phase:
\begin{equation}\label{mainfindingA}
\mathit{\Sigma}_{st}(m;\tau) = 
\mathit{\Sigma}_{eq} (m) - \frac{1+\tau}{2\tau} (1-m)^2, \quad 0<m<1. 
\end{equation}
As a function of parameters $m$ and $\tau$,  the complexity exponent $\mathit{\Sigma}_{st}$ is positive above the curve $\tau=\tau_0(m)$ in the $(m,\tau)$-plane,
\begin{equation}\label{mainfindingB}
\tau_0(m)=-\frac{1}{2}\frac{(1-m)^2}{{1-m+\ln{m}}}, \quad 0\le  m \le  1\, ,
\end{equation}
 and is negative below. Thus, this curve and the vertical line $m=1$ partition the parameter space of our model into three regions, see Fig. \ref{fig}.
If $m>1$ the nonlinear system [\ref{1}] has, on average, exactly one equilibrium and this equilibrium is stable.
This is a region of {\it absolute stability}.
If $m<1$ then the number of stable equilibria depends on the relative strength of curl-free and divergence-free components of the interaction field. If $\tau < \tau_0(m)$ then the complexity exponent $\mathit{\Sigma}_{st}$ is negative and
the probability that the system has at least one stable equilibrium is exponentially small for large $N$. This is a region of {\it absolute instability}: on average, equilibria are exponentially abundant but
only very rare realizations of the interaction field yield stable equilibria.
In contrast, if 
$\tau>\tau_0(m)$ then the complexity exponent $\mathit{\Sigma}_{st}$ is positive,
so that in this region the stable equilibria are, on average, exponentially abundant.
However,   $\mathit{\Sigma}_{st}<\mathit{\Sigma}_{eq}$ and, hence,  the stable equilibria are, on average, exponentially rare   \emph{among  all equilibria}.
This is also reflected in the fact that if $m<1$ then the probability for an equilibrium to be stable is, in the annealed approximation,  exponentially small for $N$ large regardless of the value of $\tau$: to leading order in $N$,
\begin{equation*}\label{transition}
\ln p_{st}^{(a)} 
= -N(1+\tau)(1-m)^2/(2\tau).
\end{equation*}

\begin{figure}[t!]
\centering
\includegraphics[width=.5\linewidth]{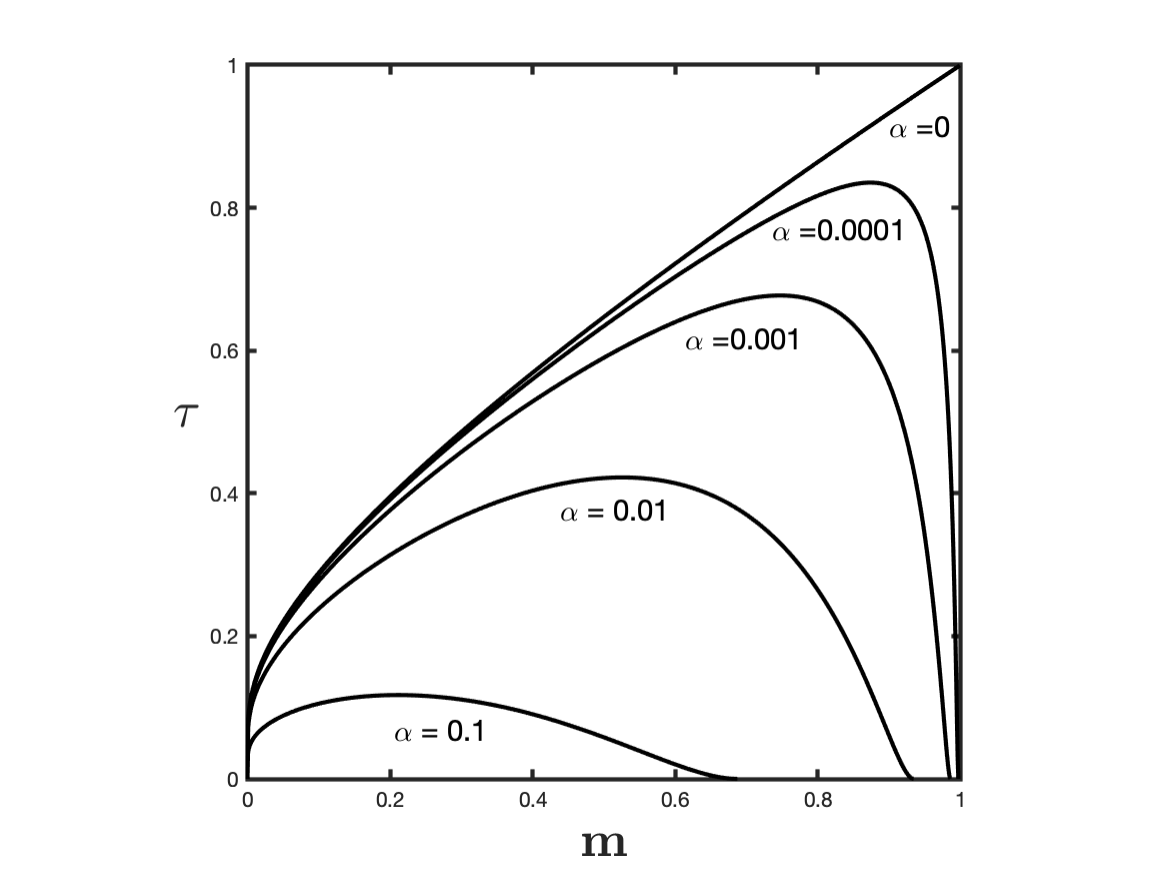}
\vspace*{-0.6ex}
\caption{The graph of $\tau_0^{(\alpha)} (m)$ as function of $m$ for $\alpha=0, 0.1, 0.01, 0.001$ and 0.0001, $\tau_0^{(0)} (m)=\tau_0 (m)$. }
\label{fig:3}
\end{figure}

One can also compute in closed form the complexity exponent associated with the $\alpha$-stable equilibria. The result of this computation is that $\mathit{\Sigma}_{st}^{(\alpha)}(m,\tau)=\mathit{\Sigma}_{eq}(m)$  for all $\alpha \ge 1/2$ and that
for all
$0\le \alpha \le  1/2$,
\[
\mathit{\Sigma}_{st}^{(\alpha)}(m,\tau) =   \begin{cases} \displaystyle{\mathit{\Sigma}_{eq} (m) -  \frac{1+\tau}{2\tau} (m_{\alpha}-m)^2},  & 0< m \le  m_{\alpha}, \\ 
  \mathit{\Sigma}_{eq} (m), & m_{\alpha}  \le  m < 1,  
   \end{cases}
\]
where $m_{\alpha}$ is the solution of equation
\begin{equation}\label{malpha}
\alpha=\frac{1}{\pi}
\big(
\arccos{m}-m\sqrt{1-m^2}
\big)
\end{equation}
for $m$.
The zero-level line of $ \mathit{\Sigma}^{(\alpha)}_{st} $, $ \tau=\tau_0^{(\alpha)}(m)$,  is given by
 \begin{equation*}\label{mainfindingBalpha}
\tau_0^{(\alpha)}(m)=\frac{(m_{\alpha}-m)^2}{-1-m_{\alpha}^2-2\ln{m}+2m m_{\alpha}}, \quad 0\le m \le m_{\alpha}\, .
\end{equation*}

\begin{figure}[t!]
\includegraphics[width=.5\linewidth]{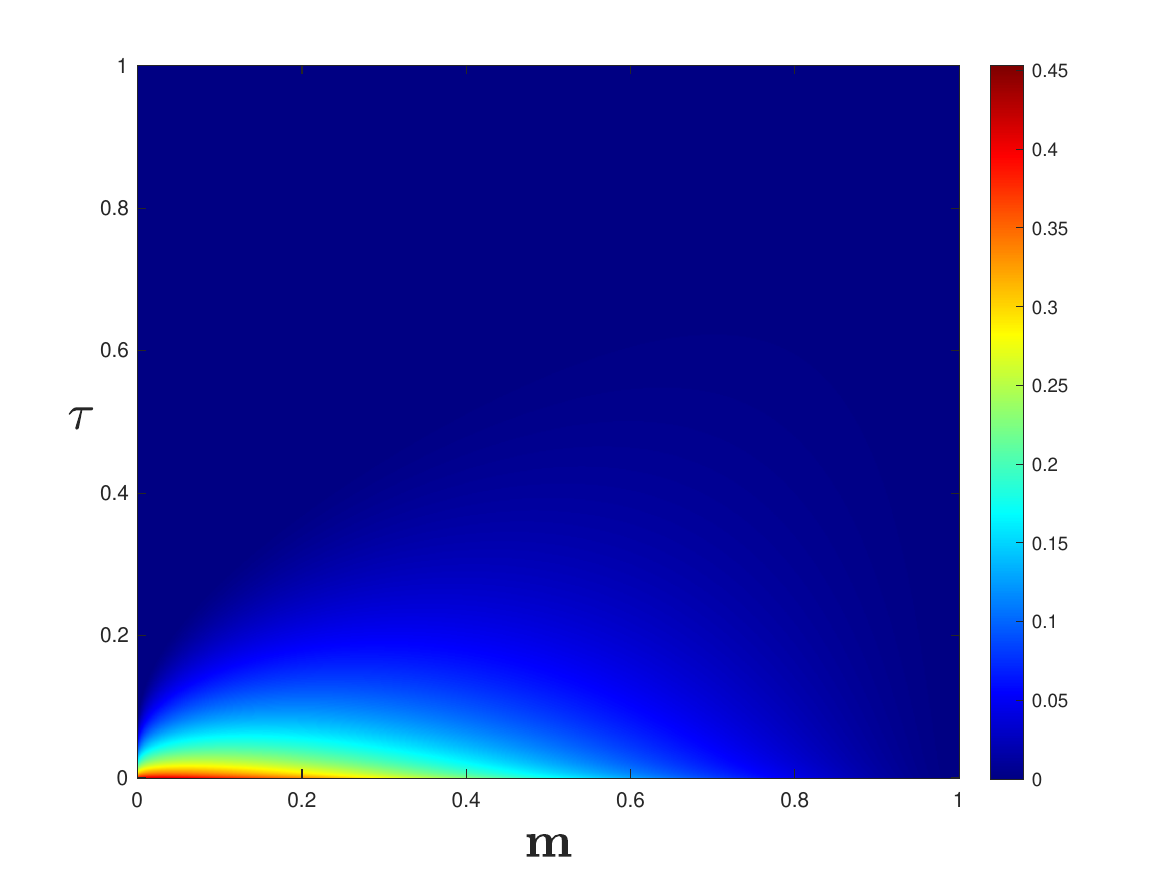}
\includegraphics[width=.5\linewidth]{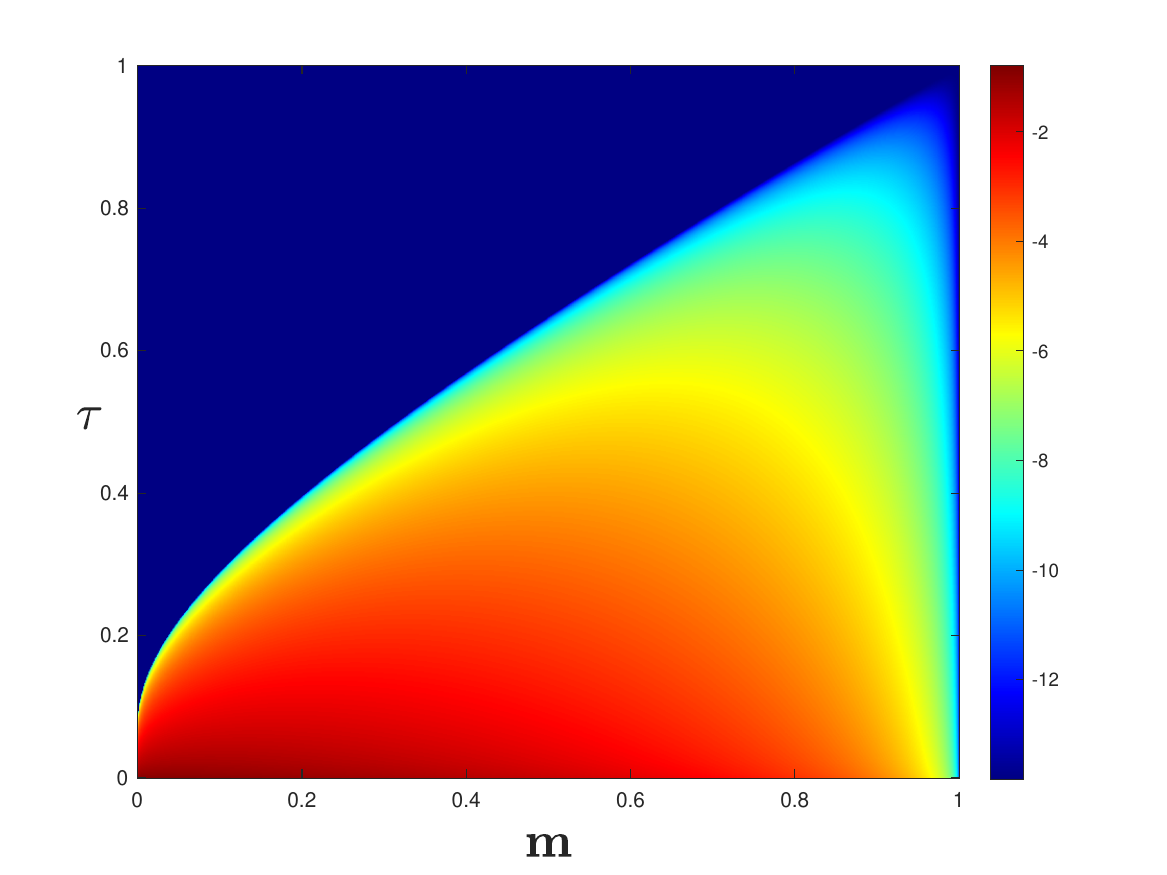}
\vspace*{-3ex}
\caption{The heat maps of $\alpha (m,\tau)$ (plot on the left-hand side) and $\ln \alpha (m,\tau)$ (plot on the right-hand side).  }
\label{fig:4}
\end{figure}

The striking feature that emerges from our analysis is the abundance of unstable equilibria with a large proportion of stable directions even far inside the absolute instability region. A quick inspection of Fig.~\ref{fig:3} leads to the conclusion that even though below the line $\tau=\tau_0 (m)$  the probability for the system to have at least one stable equilibrium is exponentially small, equilibria with a large proportion of stable directions are in abundance in this parameter range. This surprising feature can be visualized in the following way. For every point $(m,\tau)$ below the line $\tau=\tau_0(m)$ there is a unique value of $\alpha$ such that the zero-level line of $ \mathit{\Sigma}^{(\alpha)}_{st} $ passes through this point.
This mapping $(m,\tau) \to \alpha$ defines a function $\alpha (m,\tau)$ which we extend into the region above the line  $\tau=\tau_0(m)$ by setting $\alpha (m,\tau)\equiv 0$ everywhere in this region. The heat map of $\alpha (m,\tau)$, the plot on the left-hand side in Fig.~\ref{fig:4}, reveals that there is not much difference between points above  and below the critical line  $\tau=\tau_0(m)$  apart from a small 
area near $\tau=0$. For example, the zero-level lines of $ \mathit{\Sigma}^{(\alpha)}_{st} $ are barely visible (compare Fig.~\ref{fig:3} and the plot on the left-hand side in Fig.~\ref{fig:4}). One only recovers zero-level lines $\tau=\tau^{(\alpha)}(m)$ from the heat map of $\ln \alpha (m,\tau)$, see the plot on the right-hand side in Fig.~\ref{fig:4}.

\begin{figure}[t!]
\includegraphics[width=.5\linewidth]{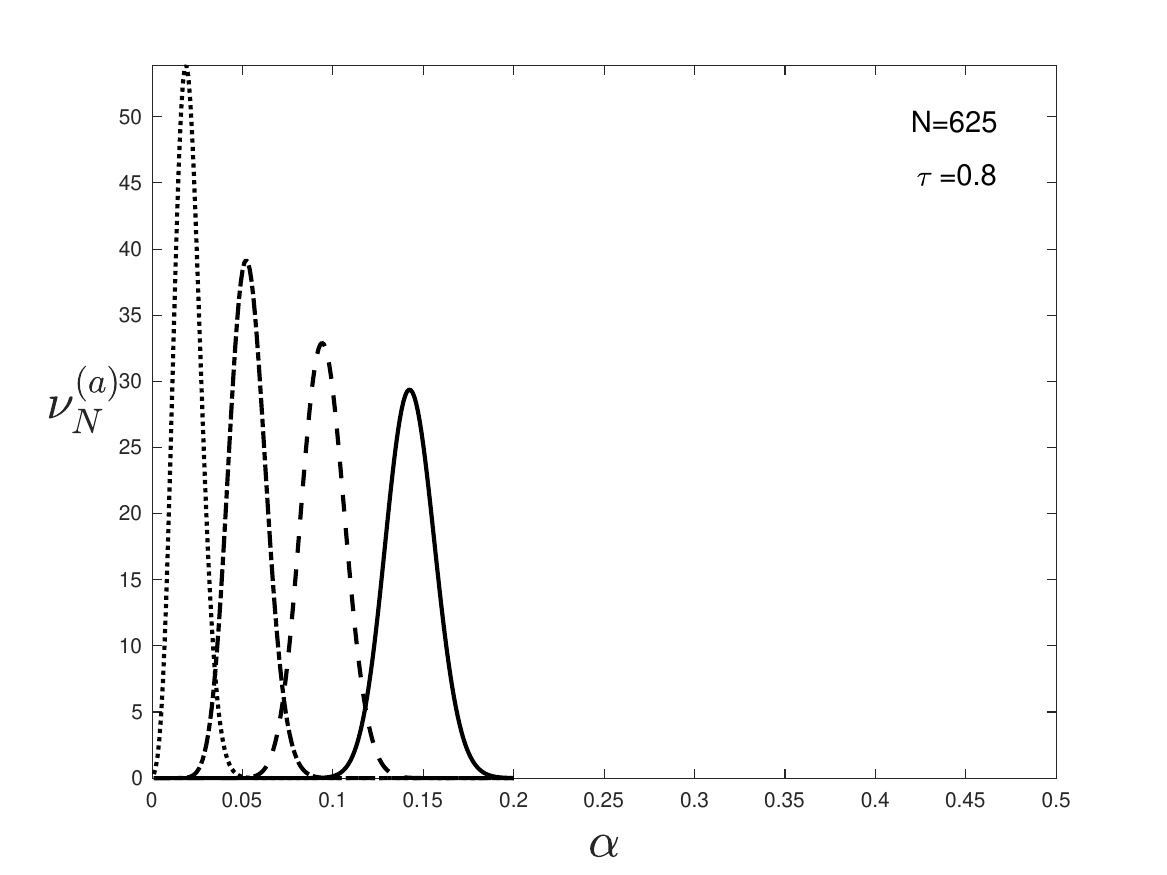}
\includegraphics[width=.5\linewidth]{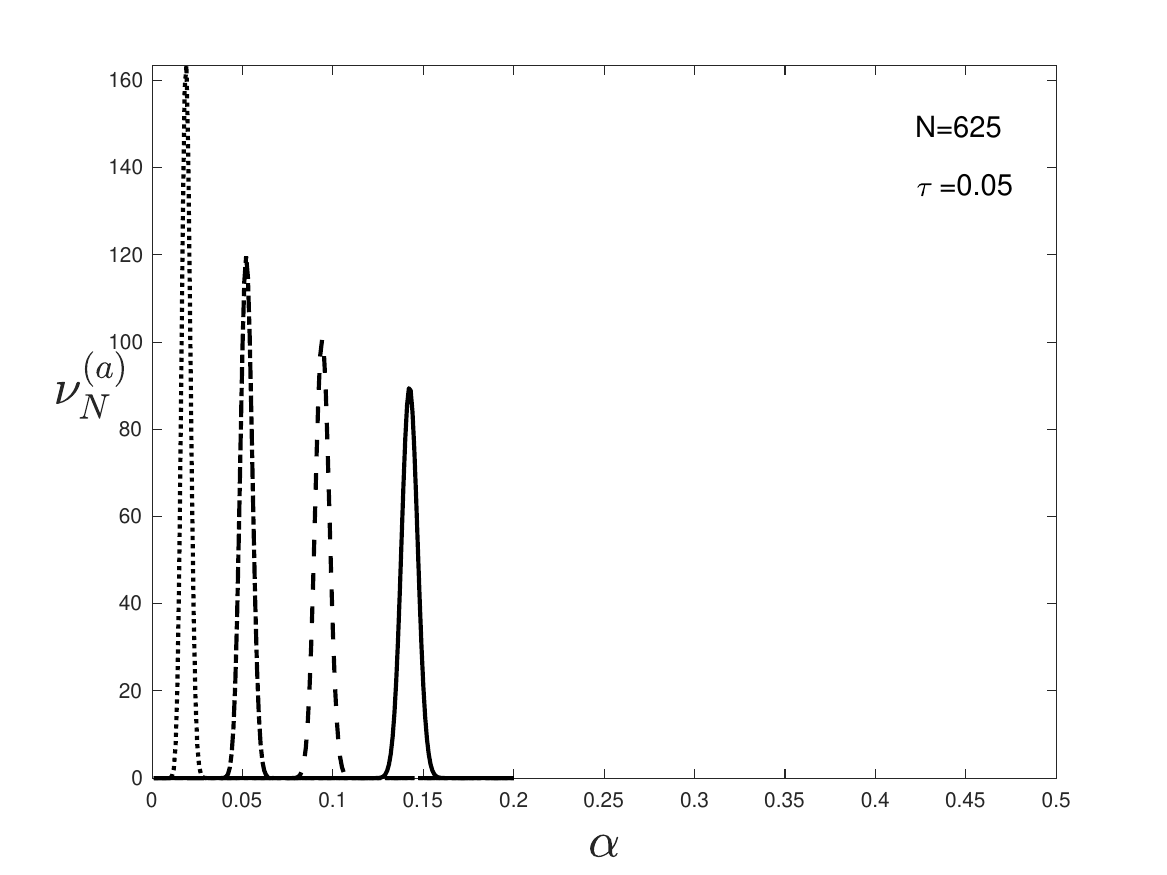}
\vspace*{-3ex}
\caption{Plot of the relative density of instability index of typical equilibria $\nu^{(a)}_N(\alpha)$ for $N=625$ and $m=0.9$ (dotted line), $m=0.8$ (dash-dotted line), $m=0.7$ (dashed line) and $m=0.6$ (solid line). $\tau=0.8$ in the plot on the left-hand side and $\tau=0.05$ in the plot on the right-hand side. Note that typical equilibria have high proportion of stable directions even when the system complexity is relatively large.}
\label{fig:5}
\end{figure}

To clarify the last point and
to get a coherent understanding of the arising picture of indices associated with equilibria
in our system it is helpful to consider the relative  {\it density} $\nu_N^{(a)}\!({\alpha})$
of $\alpha$-stable equilibria, see Eq.~\ref{denind}.
This is the probability density function of instability index $\alpha$ in the annealed approximation.  Namely,
the probability that an equilibrium drawn at random from the entire population of equilibria will
have its instability index in the interval $(\alpha_1, \alpha_2)$ is given by the integral $\int_{\alpha_1}^{\alpha_2} \nu_N^{(a)}({\alpha}) \, d\alpha$ in the annealed approximation.
 In the limit $N\gg 1$ this density 
 can be determined in closed form in the entire range of $\alpha\in [0,1]$, including the leading pre-exponential factor, see Supplementary Information. To leading order in $N$,
\begin{equation}\label{denalpha}
\nu_N^{(a)}({\alpha}) = \frac{1}{2}\sqrt{\frac{N\pi(1 +\tau) }{2 \tau (1-m^2)}}\, e^{-\frac{1+\tau}{2}\big[\! \frac{\sqrt{N}(m_{\alpha}-m)}{\sqrt{\tau}}\big]^2},
\end{equation}
where,
for any given $\alpha\in [0, 1]$,
$m_{\alpha}$ is the (unique) solution of Eq.~\ref{malpha} for $m$ in the interval $[-1,1]$.
It is apparent that in the topologically non-trivial phase $m\in(0,1)$
only equilibria with the instability indices $\alpha$ in a narrow interval of width $\sqrt{\tau/N}$ around the value $\alpha_m
=\frac{1}{\pi}[\arccos{m}-m\sqrt{1-m^2}]$,  $0\le \alpha_m \le 1/2$,  have finite density, see Fig.~\ref{fig:5}.  
The equilibria
with index $\alpha >1/2 $ have, on average, always exponentially vanishing density relative to the total number of equilibria.
This can be seen by noticing that  $m_{\alpha}$ is negative for such values of $\alpha$.

The transition from absolute stability to instability as the system complexity increases is very sharp for $N$ large. Indeed, the complexity exponent 
$\mathit{\Sigma}_{eq}(m)$ vanished quadratically at $m=1$, hence the width of the transition region scales as $N^{-\frac{1}{2}}$.
Although our methods give no access to the entire  transition region 
one can probe its left tail  by setting 
\begin{equation*}\label{trlt}
m=1-{\delta}/ \sqrt{N}, \quad 1\ll \delta \ll \sqrt{N}.
\end{equation*}
in Eqs~\ref{mainfindingA} -- \ref{denalpha}. For example, the probability for an equilibrium to have the number of its unstable directions $\kappa$ in the interval $(\gamma_1 N^{1/4}, \gamma_2N^{1/4})$ is given,  in the annealed approximation, by $\int_{\gamma_1}^{\gamma_2}\sigma (\gamma)\,  d\gamma $
where to leading order in $N$ and $\delta$
\begin{equation*}\label{trlt1}
\sigma(\gamma)\!=\!\frac{1}{N^{3/4}}\nu^{(a)}\!\Big(\frac{\gamma}{N^{3/4}}\!\Big) =
 \sqrt{\frac{\pi(1+\tau)}{16\, \tau \delta}}
e^{
-\frac{1+\tau}{2\tau}
\big[
\delta-\frac{1}{2}
\big( \frac{3\pi}{2}\gamma
\big)^{2/3}
\big]^2
 }, 
\end{equation*}
see Supplemental Information.
In particular, this means that in the left tail of the transition region the number of unstable directions of a typical equilibrium scales with $N$ as 
$N^{1/4}$.
This leads to the natural conjecture that the number of unstable directions of typical equilibria in the entire transition region is proportional to $N^{1/4}$. In the annealed approximation this conjecture was verified in ref~\cite{GK} for the pure gradient flow. 

\subsection*{Discussion}
In this paper we extend May's local stability analysis of large complex systems from the neighborhood of a single equilibrium to the entire phase space of the system. The systems which we consider are equipped with a stability feedback mechanism and the   interaction complexity is modeled by a random field of zero mean value which couples the degrees of freedom, see Eq.~\ref{1}. Our model system is, in a certain sense, 'minimal' as it has only two control parameters, see Eq.~\ref{deftau}.

The following picture then emerges from our analysis. For large values of $m$ the stability feedback mechanism prevails and, typically, large complex systems will have only one equilibrium which is stable. This is the regime of absolute stability. For non-gradient systems ($\tau<1$),  as the interaction strength increases the system undergoes a sharp transition at the critical point $m=m_C=1$ from the regime of absolute stability to the regime of absolute instability. In latter regime the system has multiple equilibria, but the probability for the system to have at least one stable equilibrium is exponentially small. However, equilibria with a large proportion of stable directions are in abundance in this regime. With the further increase of the 
interaction strength,  the system transits to the regime of relative instability which is characterized by the abundance of stable equilibria, yet the unstable equilibria dominate. The transition point $m=m_B$ from the absolute to relative instability depends on the relative strength of the stability feedback mechanism and the balance between the gradient and solenoidal components of the coupling field, see Eq.~\ref{mainfindingB}, and if the coupling is divergence-free  ($\tau=0$) then the relative stability regime does not exists at all. If the coupling is curl-free  ($\tau=1$) then, 
as the 
 interaction strength increases, the system transits from the regime of absolute stability directly to the regime of relative instability.

We expect that some 
 qualitative features revealed in the phase portrait of the present model  may be shared by other systems of randomly coupled autonomous ODE's  with large number of degrees of freedom, such as e.g. a  model of neural network consisting of randomly interconnected neural units \cite{WT}, or non-relaxational version of the spherical spin-glass model \cite{YF2016,CKDP}.
 Earlier studies, starting from the classical paper \cite{SCS1988} suggested that autonomous dynamics in the 'topologically nontrivial' regime should be predominantly chaotic, see \cite{WT,CS2018} and references therein. The absence of stable, attracting equilibria certainly corroborates this conclusion, though presence of stable periodic orbits in the phase space can not be excluded on those grounds either. The influence of the non-gradient 
component of the vector field on system dynamics needs further clarification as well. On one hand, as we discovered above any admixture of such components very efficiently eliminates all stable equilibria when entering the 'topologically non-trivial' regime. On the other hand, the results of the paper \cite{CKDP} suggest that the influence of such  non-potentiality on long-time 'aging' effects in dynamics of glassy-type models is relatively benign. This may imply that the dynamical dominance of exponentially abundant, though unstable equilibria with yet extensively many stable directions 
may be enough for 'trapping' the system dynamics for a long time in the vicinity of such equilibria, thus inducing aging phenomena similar to the gradient descent case  \cite{Kurchan_1996,CK1993}. 

As one of main outstanding challenges one must mention obtaining statistical characteristics of ${\cal N}_{eq}$ and ${\cal N}_{st}$ beyond their mean values. As it is known that ''quenched'' and ''annealed'' complexity of minima may not coincide in some models of random landscapes, see e.g. \cite{Auf1,Subag2017}, one may expect that such a calculation may lead to a further refinement of
the picture of transition lines presented in our paper for certain classes of random functions $V(\vec{x})$ and $A_{ij}(\vec{x})$. 
Recent progress in purely gradient case is encouraging, see \cite{Subag2017,Ros2018} and hopefully can be extended to the general case.
Apart from that,  studying dynamical equilibria in ecological models with species-dependent relaxation rates, or
 structured interspecies interactions \cite{Biroli2018}, and  investigating similar questions in other related models with non-gradient dynamics, see e.g. \cite{Ipsen2017,IpsenForr2018,FSK2017}  looks  promising.
 
 In the next Section we will outline our methods. Whilst in the purely gradient case $\langle {\cal N}_{st}^{(\alpha)}\rangle $, the expected value of the number of $\alpha$-stable equilibria, can be related to the probability distribution of the $(\alpha N+1)$-st top eigenvalue of the Gaussian Orthogonal Ensemble  paving way to a precise and mathematically rigorous asymptotic analysis of $\langle {\cal N}_{st}^{(\alpha)}\rangle $, we are not aware of an analogous relation for the non-gradient systems ($\tau<1$). In fact, the probability distribution of the $(\alpha N+1)$-st largest real part of eigenvalues of real random matrices is unknown, even for $\alpha=0$ and finding it presents a very challenging and highly nontrivial mathematical problem. Our approach in the non-gradient case utilizes theory of large deviations for eigenvalues of random matrices. Once consequence of this is that with the exception of the density of unstable directions $\nu_N^{(a)}({\alpha}) $ we can only obtain the complexity exponents but not the pre-exponential factors (and hence cannot access the transition region around the instability threshold at $m=1$). Another is that the probability of large deviations of the $(\alpha N+1)$-st largest real part of eigenvalues is unknown at the required precision level and is left conjectured. Validating this conjecture and giving full mathematical justification of our formal asymptotic analysis remains an outstanding probabilistic problem.  
\matmethods{
Our analysis of stability of equilibria in model [\ref{1}]
is based on a Kac-Rice integral representation of the average value of linear statistics of equilibria in terms of a random matrix average [\ref{LS1}] and a subsequent use of random matrix techniques. 
Technical details of our calculations can be found in Supplemental Information. Here we focus on the main ideas and the assumptions we have used.

Suppose the test function $\mathit{\Psi}$ in Eq.~\ref{LS1} is given by $\mathit{\Psi} (\vec{x})=\psi (J(\vec{x}))$ where $J(\vec{x})=(J_{ij})$ is the Jacobian of the interaction  field $\vec{f}(\vec{x}) $.
Then applying the Kac-Rice formula, see, e.g.,  refs \cite{AT2007} and \cite{AW2009},
\begin{equation}\label{KC_int}
\langle L[\mathit{\Psi}]  \rangle_{\!\vec{f}}\! =\!\!\int_{\mathbb{R}^N}\!\!\!\big\langle   \psi (J_{ij})  |\det ( J_{ij} \!-\!\mu \delta_{ij})| 
\prod_j  \delta ( f_j(\vec{x})\!-\!\mu x_j)  \big\rangle_{\!\!\vec{f}} \, d x.
\end{equation}
Under our assumptions on the law of distribution of $\vec{f}(\vec{x})$, the integrand  
factorizes into the product of $\langle  \psi (J_{ij})  |\det ( J_{ij} -\mu \delta_{ij})|\rangle_{\!\vec{f}} $ and $\langle  \prod_j  \delta ( f_j(\vec{x})-\mu x_j)\rangle_{\!\vec{f}} $, and the integral can easily be evaluated,  see ref. \cite{FyoKhor2016}. Since the matrix-valued field $J(\vec{x})$ is homogenous, the first factor is independent of $\vec{x}$, and the second factor, 
when integrated over $\vec{x}$, yields $1/\mu^N$, hence Eq.~\ref{LS1} which gives $\langle L[\mathit{\Psi}]  \rangle_{\!\vec{f}}$ in terms of a random matrix average.

The underlying random matrix distribution can be found by differentiating Eqs \ref{3a}--\ref{3b}. This gives the the covariance function of the matrix entries of $J$, and since $J$ is Gaussian, also its distribution. The result of this calculation is that to leading order in $N$, 
\begin{equation*}
J  \,{\buildrel d \over =}\, \sigma \sqrt{N} (X-\xi I) \quad (N\gg 1)
 \end{equation*}
where $ \sigma=2\sqrt{v^2+a^2}$ and the matrix $X$ and scalar $\xi$ are independent Gaussians. The scalar $\xi$ has mean value zero and variance $\tau/N$ and the matrix distribution of $X$ is given by 
\begin{equation*}\label{X}
\langle X_{ij}\rangle=0, \quad \left\langle X_{ij}X_{nm} \right\rangle=N^{-1}(\delta_{in}\delta_{jm}+  \tau \delta_{jn}\delta_{im})\, ,
\end{equation*}
The ensemble of matrices $X$ interpolates between the Gaussian Orthogonal Ensemble of real symmetric matrices ($\tau=1$) and real Ginibre ensemble of fully asymmetric matrices ($\tau=0$) and is known as the real elliptic ensemble, see refs \cite{ForNag2008,KS} for details.

Thus, to leading order in the limit $N\gg 1$,
\begin{equation}\label{L[Phi]}
\langle L[\mathit{\Psi}]  \rangle_{\mathbf{f}} =m^{-N}  \langle \psi (\sigma \sqrt{N} (X-\xi I)) |\det (X-(\xi+m) I) |\rangle_{X,\xi}\, ,
\end{equation}
where the average on the right-hand side is over $\xi$ and $X$. By setting here
$\psi=\mathit{\Theta}(\mu-x_{max}(J))$ in Eq.~\ref{L[Phi]} one obtains the average number of stable equilibria: 
 \begin{equation}\label{Nstmain}
\left\langle{\cal N}_{st}\right\rangle\!=\! \frac{1}{m^N} \int_{-\infty}^{\infty} \mathcal{D}_N(x)\, e^{-\frac{N(x-m)^2}{2\tau}} \, \frac{ dx}{\sqrt{2\pi\tau/N}}
\end{equation}
with
$\mathcal{D}_N(x)= \!\! \langle \mathit{\Theta} (x-x_{max}) \left|\det (X-xI)\right|  \rangle_X $,
where $x_{max}$   
is the largest real part of the eigenvalues 
of $X$.
The Elliptic Law, see refs~\cite{G86, Nguyen14} asserts that in the limit $N\gg 1$ the eigenvalues of $X$ are uniformly distributed in the domain
${x^2}/{(1+\tau)^2}+{y^2}/{(1-\tau)^2}\le 1$ in the complex plane $z=x+iy$. Correspondingly, the asymptotic behavior  of
$\mathcal{D}_N(x)$ will depend on position of $x$ relative to this elliptic domain.

If $x>1+\tau$ then typical realizations of $X$ will have all of its eigenvalues located left to the vertical line $\Re z = x$ and the constraint $x_{max}<x$ in the definition of  $\mathcal{D}_N(x)$ is not satisfied only in a rare event. It can be shown, see Supplemental Information, that the probability of such an event is exponentially small, and, consequently, to leading order in $N$,
\begin{equation}\label{D_N_+}
\ln \mathcal{D}_N(x) =\ln \langle \left|\det \left(X-xI\right)\right| \rangle_X = {N\mathit{\Phi}(x; d\mu_{eq})}, \quad x>1+\tau.
\end{equation}
Here, $d\mu_{eq}$ is the limiting elliptic eigenvalue distribution of $X$ and $\mathit{\Phi}(x; d\mu_{eq})$ its log-potential,
\[
\mathit{\Phi} (x; d\mu)=\int \ln |z-x| \, d\mu (z)\, .
\]

If $x<1+\tau$ then in
this case, typical realizations of $X$ will have a macroscopic number of eigenvalues located right of the vertical line $\Re z=x$ and only in very rare realizations of $X$ the condition $x_{max}<x$ is satisfied. It follows from large deviation theory for random matrices that in the limit $N\gg 1$ all such realizations have the same eigenvalue distribution $d\mu_x$ which is the minimizer of the large deviation rate functional
\begin{eqnarray} \label{LDPel}
\lefteqn{{\cal J}_{\tau}[d \mu ]= \frac{1}{2}\int_{\mathbb{C}} \left[ \frac{(\Re z)^2}{1+\tau} +
\frac{(\Im z)^2}{1-\tau} \right]\, d\mu (z) }\\
& & \nonumber
\hspace*{20ex}
- \frac{1}{2} \int_{\mathbb{C}}\int_{\mathbb{C}}\log |z-w|\,  d\mu (z)
d\mu (w) - \frac{3}{8}
\end{eqnarray}
on the set of all probability distributions in complex plane  whose support lies left of the vertical line $\Re z=x$ and which are symmetric with respect to reflection in the real line. Also, to leading order,
\begin{equation}\label{LDP3}
\ln \Prob( x_{max } <x) = - N^2K_{\tau}(x), \quad K_{\tau}(x) = \mathcal{J}_{\tau}[d\mu_x ]\, .
\end{equation}
Correspondingly, see Supplemental Information, $\mathcal{D}_N(x)$ factorizes: 
\begin{equation}\label{factorisation}
\mathcal{D}_N(x) = e^{N\mathit{\Phi}(x; d\mu_x)+o(N)} \Prob \{ x_{max } <x\}, \quad x<1+\tau\, .
\end{equation}
Determining the minimizer $\mu_x$ of the large deviations rate functional 
in closed form is a highly nontrivial exercise in potential theory, which, at present, is only solved in the special case $\tau=1$ \cite{BDG2001,DM_SP}, and is partly characterized for $\tau=0$  in \cite{ASZ2014}.
Fortunately, for our purposes, the exact form of the minimizer $\mu_x$ is not needed, apart from the following continuity property of the log-potential:
\begin{equation}\label{cont}
\lim_{x\to1+\tau-0}\mathit{\Phi} (x; d\mu_x)=\lim_{x\to1+\tau+0}\ \mathit{\Phi} (x; d\mu_{eq})\, .
\end{equation}

Eqs~\ref{D_N_+} -- \ref{factorisation} suggest that the integral in Eq.~\ref{Nstmain} can be asymptotically evaluated for $N\gg 1$ by the Laplace method. Such an evaluation is indeed possible, and it leads to Eq.~\ref{mainfindingA}, but it involves a subtle step which we should mention here, for details see Supplemental Information. It can be shown that the main contribution to this integral is coming from a small neighborhood of $x=1+\tau$. But then, since $K_{\tau}(x)$ vanishes as $x$ approaches $1+\tau$,  next-to-leading order corrections to Eq.~\ref{LDP3} cannot be ignored. In other words,  for our goal of evaluating  the integral in Eq.~\ref{Nstmain} the precision of Eq.~\ref{LDP3}  is not sufficient. What is actually needed is a sharper large deviation principle which includes the next sub-leading term in the exponential. We conjecture that this term is of order $N$:
\begin{equation}\label{LDP4}
\Prob \{ x_{max} <x \} = e^{-N^2K_{\tau}(x)-NT_{\tau}(x)+o(N)}\, ,   \quad x<1+\tau.
\end{equation}
Our conjecture is based on a similar sharper large deviation principle for the largest eigenvalue of Gaussian Hermitian and real symmetric matrices in the framework of a powerful, albeit heuristic version of the Large Deviation Theory for random matrices  known as the 'Coulomb gas' method,
see calculations in, e.g.,  ref.~\cite{Borot_2011} and, closer to our context, in Appendix C of ref.~\cite{FyoWi07}.
Although similar heuristic justifications for the validity of Eq.~\ref{LDP4} can be provided for our case as well,
a rigorous verification of such sharp large deviation principle 
is an open challenging problem, for a related work see ref.~ \cite{SL}. 

The average number of $\alpha$-stable equilibria, $\langle{\cal N}_{st}^{(\alpha)}\rangle$,
and the density of the number of unstable directions, 
$\nu_N^{(a)}(\alpha)$, is evaluated along similar lines, for details see Supplemental Information.
}

\showmatmethods{} 

\acknow{
We are indebted to J.-P. Bouchaud who after reading \cite{FyoKhor2016} informally conjectured the existence of another transition  below $m_C=1$
and encouraged two of us to investigate the stability index of equilibria as well as to look for the phase boundary $\tau=\tau_0(m)$ in the $(m, \tau)$ plane. 
Also, we thank  J. Grela, S. Sodin and O. Zeitouni for their constructive critique of earlier versions of this manuscript.  
 }

\showacknow{} 
\bibliography{BFK}

\end{document}





%
%
%

\maketitle
\SItext

\section{Introduction}
These notes provide details of our calculation of $\left\langle{\cal N}_{st}\right\rangle$ and $\langle{\cal N}_{st}^{(\alpha)}\rangle$, the average numbers of stable and $\alpha$-stable equilibria, 
in the `minimal' nonlinear model for large complex systems,
\begin{equation}\label{A:1}
\dot{\vec{x}}=- \mu \vec{x}+ \vec{f}(\vec{x}),  \quad  \vec{x} \in \mathbb{R}^N\, , \quad  \mu >0\, ,
\end{equation}
and, also, the relative density of $\alpha$-stable equilibria,
 \begin{equation}\label{denind}
\nu^{(a)}_N(\alpha)=\frac{1}{\langle \mathcal{N}_{eq}\rangle}
\frac{d}{d\alpha}\langle \mathcal{N}^{(\alpha)}_{st}\rangle\, .
\end{equation}
The interaction field $\vec{f}(\vec{x})$ is a zero mean homogeneous isotropic Gaussian random field with smooth realisations. The `minimal' nonlinear model has two control parameters: the scaled relaxation strength  $m$,
\[
m=\frac{\mu}{2\sqrt{N(v^2+a^2)}}
\]
and the non-potentiality parameter  $\tau\in[0,1]$,
\[
\tau=\frac{v^2}{a^2+v^2}\, ,
\]
see Section Model Assumptions for the definitions of parameters $a$ and $v$. In the context of large complex systems, the  
parameter $m$ eventually reflects the complexity of the system's phase portrait: the lower is the value of $m$ the more complex is the system's phase portrait due to presence of large number of different equilibria. As to the parameter $\tau$, it mainly controls the balance between longitudinal and transversal components of the interaction field. If $\tau=1$ then $\vec{f}(\vec{x})$ is curl free and if $\tau=0$ then $\vec{f}(\vec{x})$ is divergence free.

The value $m=1$ is the instability threshold in the model, see ref.~\cite{FyoKhor2016}.  If $m>1$ then in the limit $N\gg 1$ the nonlinear system Eq.~\ref{A:1} has on average only one equilibrium which is stable. In contrast,  if $m<1$ then, on average, the system has exponentially many equilibria and its phase space is topologically non-trivial. In these notes, unless explicitly stated otherwise, we always assume
\[
\tcboxmath{
0<m<1\, , \quad 0\le \tau < 1\, .
}
\]
The boundary case of $\tau=1$ corresponds to purely gradient descent dynamics. In this case $-\mu \vec{x} + \vec{f(x)} =\nabla L (\vec{x})$ for some Lyapunov function $L (\vec{x})$ and counting stable equilibria is equivalent to counting local minima on the surface of $L (\vec{x})$ which was performed previously using a variety of different methods.
Although in our setup the case of $\tau=1$ is singular, see Eq.~\ref{JPDGin}, with suitable modifications 
our calculation goes through in this case too and reproduces the expressions for $\left\langle{\cal N}_{st}\right\rangle$ and $\langle{\cal N}_{st}^{(\alpha)}\rangle$ obtained in \cite{BD07,FyoWi07,FyoNad2012,Auf1}. However, in the interest of uniformity of the presentation we avoid here the singular case restricting ourselves to the parameter range $0\le \tau < 1$.

\medskip

The emphasis in these notes is more on computation rather than on trying to provide full mathematical justification for each step of our calculations. In some places, a different approach can be used to arrive at the same result which is more mathematically rigorous but less intuitive, e.g., Eq.~\ref{LDP_B}  in Section \ref{subsection_B}, in some places tedious calculations can fill the gap, e.g., dealing with the log-singularities in Section \ref{subsection_E} and in some places the required results are left conjectured, e.g., finding the next-order corrections in the large deviation  principle for the empirical eigenvalue counting measure, Section \ref{subsection_F}. Validating this conjectures and giving full mathematical justification of our formal asymptotic analysis
remains an outstanding probabilistic problem. 

Throughout these notes we shall use the notation $x_{max}(J)$ to denote the largest real part of the eigenvalues of matrix $J$ and $\mathit{\Theta (x)}$ will stand for the Heaviside step function,
\begin{eqnarray}\label{Heav}
\mathit{\Theta} (x) =
\begin{cases}
1, & \mbox{if } x>0, \\
0, & \mbox{if } x<0 .
\end{cases}
\end{eqnarray}
Both counting functions ${\cal N}_{st}$ and ${\cal N}_{st}^{(\alpha)}$  are examples of linear statistics of equilibria
\[
L[\psi] = \sum_{\vec{x}_*} \psi (J_*)\, .
\]
Here, the sum is over all equilibria of Eq.~\ref{A:1} and $\psi $ is a test function of matrix argument. The notation $J_*$ stands for the Jacobian matrix of the interaction field $\vec{f}(\vec{x})$ at equilibrium $\vec{x}_*$,
\[
J_* = \left. \left( \frac{\partial f_j}{\partial x_k} \right)\right|_{\vec{x}=\vec{x}_*}\, .
\]
Choosing the test function in the form $\psi _{st}(J)= \mathit{\Theta} (\mu-x_{max} (J))$, one obtains  $ L[\psi _{st}]={\cal N}_{st}$ and if
\[
\psi _{st}^{(\alpha)}(J)= \mathit{\Theta} \Big(\alpha-\int_{\mu}^{+\infty} \!\!\!dx \!\int_{-\infty}^{+\infty} \!\!\!dy\,\, \rho(x, y; J)\Big), \quad \rho(x, y; J)=\frac{1}{N} \sum_{j=1}^N \delta (x -\Re z_j(J)) \delta (y -\Im z_j(J))
\]
were $z_j (J)$ are the eigenvalues of $J$,  then $L[\psi _{st}^{(\alpha}]={\cal N}_{st}^{(\alpha)}$.

\medskip

The equilibria of Eq.\ref{A:1} are { \color{blue} defined as} roots of the equation $0=- \mu \vec{x}+ \vec{f}(\vec{x})$. An application of the Kac-Rice formula for counting zeros of random functions then yields the expected value of linear statistics of equilibria in terms of a random matrix average:
\begin{equation}\label{LS1}
\langle L[\psi]  \rangle_{\vec{f}} = \mu^{-N} \langle \psi (J) \, |\!\det(-\mu I +J )|\rangle_J\, .
\end{equation}
Here, the angle brackets on the left-hand side stand for averaging over realisations of the interaction field $\vec{f}(\vec{x})$, and the angle brackets on the right-hand side stand for averaging over the distribution of the Jacobian matrix $J=(\partial f_j/\partial x_k)$. The latter does not depend of $\vec{x}$ because of the homogenuity of $\vec{f}(\vec{x})$.   An outline of our derivation of Eq.~\ref{LS1} is given in Section Materials and Methods and technical details can be found in ref.~\cite{FyoKhor2016}.

The probability distribution of the Jacobian matrix $J$ can be found by differentiating the covariance function of the random field $\vec{f}(\vec{x})$. The result of this calculation is that to leading order in $N$, 
\begin{equation}\label{F_x}
J  \,{\buildrel d \over =}\, \sigma \sqrt{N} (X-\xi I) \quad (N\gg 1)
 \end{equation}
where $ \sigma=2\sqrt{v^2+a^2}$ and the matrix $X$ and scalar $\xi$ are independent Gaussians. The scalar $\xi$ has mean value zero and variance $\tau/N$ and the matrix entires of $X$ have zero mean and covariance
\begin{equation*}\label{X}
 \left\langle X_{ij}X_{nm} \right\rangle=N^{-1}(\delta_{in}\delta_{jm}+  \tau \delta_{jn}\delta_{im})\, ,
\end{equation*}
This ensemble of random matrices interpolates between the Gaussian Orthogonal Ensemble of real symmetric matrices ($\tau=1$) and real Ginibre ensemble of fully asymmetric matrices ($\tau=0$) and is known as the real elliptic ensemble ($\rGin(\tau, N)$), see refs \cite{ForNag2008,KS} for details. Alternatively, the real elliptic ensemble  can be defined by the probability (ensemble) distribution $dP(X)={\cal P}(X)\prod_{i,j=1}^N dX_{ij}$ on the space of real $N\times N$ matrices with density
\begin{equation}\label{JPDGin}
\tcboxmath{
  {\cal P}(X)\propto  \exp \left[
  {-\frac{N}{2(1-\tau^2)} \sum_{j,k=1}^N (X_{jk}^2-\tau X_{jk} X_{kj})}
  					\right]
=
 \exp \left[
{-\frac{N}{2(1-\tau^2)}\Tr  (X\,X^T-\tau X^2 )}
\right] \, .
}
\end{equation}

Eqs~\ref{LS1} -- \ref{F_x} imply that to leading order
\begin{equation}\label{L[Phi]}
\langle L[\psi]  \rangle_{\vec{f}} =m^{-N}  \langle \psi (\sigma \sqrt{N} (X-\xi I)) |\det (X-(\xi+m) I) |\rangle_{X,\xi}\, ,
\end{equation}
where the average on the right-hand side is over $\xi$ and $X$. By setting here
$\psi=\psi _{st}$  one obtains, after some straightforward manipulations, the average number of stable equilibria. To leading order in $N$,
\begin{equation}\label{A:Nstmain}
\tcboxmath{
\big\langle{\cal N}_{st}\big\rangle=\frac{1}{m^N} \int_{-\infty}^{\infty} \!  \mathcal{D}_N(x)\,  \frac{e^{-\frac{N(x-m)^2}{2\tau}} }{\sqrt{2\pi\tau/N}}\, dx \,  ,  \quad \quad  \text{where}  \quad \mathcal{D}_N(x)= \big\langle \mathit{\Theta} (x-x_{max}(X)) \left|\det (X-xI)\right|  \big\rangle_X   \, .
}
\end{equation}
Similarly, by setting $\psi=\psi _{st}^{(\alpha)}$ in Eq.~\ref{L[Phi]} one obtains
\begin{equation}\label{A:Nstmain_intro}
\tcboxmath{
\big\langle{\cal N}_{st}^{(\alpha)}\big\rangle=\frac{1}{m^N} \int_{-\infty}^{\infty} \!\!  \mathcal{D}_N^{(\alpha)}(x)\, \frac{e^{-\frac{N(x-m)^2}{2\tau}} }{\sqrt{2\pi\tau/N}}\, dx \,  \,  ,  \text{where} \,
\mathcal{D}_N^{(\alpha)}\!(x)\!=  \Big\langle \mathit{\Theta} \Big(\alpha\!-\!\int_x^{+\infty} \!\!\!\! dt \!\int_{-\infty}^{+\infty} \!\!\!\! ds\,  \rho_N(t, s; X)\Big) \left|\det (X-xI)\right|\Big\rangle_X   \, .
}
\end{equation}
If $\alpha=0$ then a quick reflection on the expression for  $\mathcal{D}_N^{(\alpha)}$ above leads to the conclusion that in this case $\mathcal{D}_N^{(\alpha)}=\mathcal{D}_N$. In fact, $\mathcal{D}_N^{(\alpha)}$ can be written in such a way where this relation is apparent. To this end, let us order the eigenvalues $z_1, \ldots, z_N$  of $X$ by their real parts $x_1, \ldots, x_N$ so that\footnote{In the real elliptic ensemble, with probability one all eigenvalues are distinct and, therefore, this labelling (ordering) arrangement is consistent.}
\[
x_{max}=x_1\ge x_2\ge x_3 \ge  \ldots \ge x_{N}\, .
\]
Then
\begin{equation}\label{NstmainA_1}
\mathcal{D}^{(\alpha)}_N(x) =   \big\langle \mathit{\Theta} (x-x_{\alpha N +1}(X))\, |\det (X-xI)| \big\rangle_X  \, ,
\end{equation}
and it is obvious that $\mathcal{D}^{(0)}_N(x)=\mathcal{D}_N(x)$. The function $\mathcal{D}^{(\alpha)}_N(x)$ can be written in yet another useful form. First, we need to introduce more notations.

Throughout these notes we will use the notation $d\mu_N(z)$ to denote the empirical eigenvalue distribution of $N\times N$  matrix $X$ in the complex plane $z=x+iy$,
\begin{equation}\label{dmu_N}
d\mu_N(z)=\rho_N(x,y)dxdy\, , \quad\quad \rho_N(x,y)= \frac{1}{N}\,  \sum_{j=1}^N \delta(x -x_j)\delta(y-y_j),
\end{equation}
where $z_j=x_j +iy_j$, $j=1, \ldots, N$, are the eigenvalues of $X$.  
Then if $\mathbb{H}_x$ is the complex half-plane on the right of the vertical line $\Re z=x$,
\[
\mathbb{H}_x=\{z\in \mathbb{C}: \, \Re z \ge x  \}\, ,
 \]
then
\[
 \int_x^{+\infty} \!\!\!\!\!\! dt \!\int_{-\infty}^{+\infty} \!\!\!\!\!\! ds\,  \rho_N(t,s) = \int_{\mathbb{H}_x} d\mu_N(z)= \mu_N(\mathbb{H}_x) \, .
\]
In these notations,
\begin{equation}\label{NstmainA_2}
\mathcal{D}^{(\alpha)}_N(x) =   \big\langle \mathit{\Theta} (\alpha -\mu_N(\mathbb{H}_x) \, |\det (X-xI)| \big\rangle_X  \, .
\end{equation}
For the purpose of our derivations, it is convenient to write the random matrix averages in Eqs~\ref{A:Nstmain}--\ref{NstmainA_2} in terms of the empirical eigenvalue distribution of $X$ and its log-potential  $\Phi (x; d\mu)$,
\begin{equation}\label{Phi_N}
\tcboxmath{
\mathit{\Phi} (x; \mu) =  \int_{\mathbb{C}} \ln |x-w| d\mu (w) =  \int_{-\infty}^{+\infty}\!\!  \int_{-\infty}^{+\infty}  \ln |x-(s+it)| \, \rho (s,t)\, ds dt   \, .
}
\end{equation}
We have
\begin{equation}
\label{b_xmax}
\tcboxmath{
\mathcal{D}_N(x) =  \big\langle \mathit{\Theta} \big(x-x_{max}(X)\big)\, e^{N\mathit{\Phi} (x; \mu_N)}   \big\rangle_X
}
\end{equation}
and
\begin{equation}
\label{D_alpha}
\tcboxmath{
\mathcal{D}^{(\alpha)}_N(x)   = \big\langle \mathit{\Theta} \big(\alpha -\mu_N(\mathbb{H}_x)\big) \, e^{N\mathit{\Phi} (x; \mu_N)} \big\rangle_X  \, .
}
\end{equation}
Although the empirical eigenvalue distribution $d\mu_N(z)$ depends on the realization of $X$, the Elliptic Law, see refs~\cite{G86, Nguyen14} ,  asserts that  $d\mu_N(z)$  converges to a non-random distribution $d\mu_{eq}(z)$ in the limit $N\gg 1$,
\begin{equation}\label{dmu_eq}
\tcboxmath{
d\mu_{eq}(z)=\rho_{eq}(x,y)dxdy, \quad\quad \rho_{eq}(x,y)=\left\{\begin{array}{cl} \frac{1}{\pi(1-\tau^2)}, & \mbox{if } \frac{x^2}{(1+\tau)^2}+\frac{y^2}{(1-\tau)^2} \le 1\, , \\[1ex] 0, & \mbox{otherwise. }
\end{array}\right.
}
\end{equation}
In other words, the Elliptic Law asserts that in the limit $N\gg 1$ the empirical eigenvalue distribution $d\mu_N$ in the real elliptic ensemble is, for typical realizations of $X$, close to $d\mu_{eq}$ and one expects that $x_{max}$ is close to $1+\tau$.  
For the purpose of our derivations we need a handle on the probability of large deviations of $x_{\max}$ from $1+\tau$ and $\mu_N$ from the Elliptic Law. This is developed in  Sections \ref{subsection_B} and \ref{subsection_C}.

\section{Right tail of the probability distribution of $x_{max}$ in the real elliptic ensemble}
\label{subsection_B}

Consider $N\times N$ matrices $X$ drawn from the real elliptic ensemble [\ref{JPDGin}] in the limit $N\gg 1$. To leading order, the eigenvalues of $X$ are uniformly distributed in the elliptic domain
\[
 \frac{x^2}{(1+\tau)^2}+\frac{y^2}{(1-\tau)^2} \le 1.
\]
Note that $x=1+\tau$ is the right-most point of the limiting elliptic eigenvalue distribution. Therefore, if $x> 1+\tau$ then the probability to find an eigenvalue of $X$ to the right of the vertical line $\Re z>x$ must be small for $N$ large. In this section we quantify this statement following a method of Forrester who addressed a similar question for Hermitian matrices in ref.~\cite{Forr12}. This method exploits asymptotic independence of eigenvalues on the global scale which holds generically for random matrix ensembles. One manifestation of the asymptotic independence is the factorization of the eigenvalue correlation functions
\[
R_k(z_1,z_2,\ldots,z_k) = \frac{N!}{(N-k)!} \int P (z_1,\ldots, z_N) \prod_{j=k+1}^N\,d^2z_{j}\, ,
\]
where $P (z_1,\ldots, z_N)$ is the joint probability density of eigenvalues. Namely, if the values of $z_j$, $j=1, \ldots, k$, are all distinct then the eigenvalue correlation functions asymptotically factorize\footnote{Nontrivial correlations between eigenvalues in the real elliptic ensemble arise on the local scale when $|z_i-z_j| \propto  1/\sqrt{N}$, see   e.g., refs~\cite{ForNag2008} and \cite{KS}.}: 
\begin{equation}\label{factorisationR}
R_k(z_1,z_2,\ldots, z_k) \sim  \prod_{j=1}^k R_1(z_j)\quad \quad (N\gg 1)\, .
\end{equation}
The one-point correlation function $R_1(z)$ is proportional to the mean density of the eigenvalue distribution [\ref{dmu_N}],
\[
R_1(z)=  N \left\langle \rho_N(x,y)\right\rangle_X,
\]
so that integrating $R_1(z)$ over a domain $D$ in the complex plane yields the average number of eigenvalues of $X$ in this domain. 

\medskip
Denote the probability density function of $x_{max} (X)$ by $p_N(x)$, so that 
\begin{equation}\label{bbb}
\Prob\{x_{max}> x \} = \int_x^{+\infty} p_N(x) dx\, . 
\end{equation}
Eq.~\ref{factorisationR} can be used to approximate $p_N(x)$ for every $x>1+\tau$ in terms of
of an integral of $R_1(z)$ over $\Im z$. As in the real elliptic ensemble $R_1(z)$ is known in closed form for every finite $N$, see ref.~\cite{ForNag2008}, this gives a handle on an asymptotic evaluation of 
$\Prob\{x_{max}> x \} $ for $x>1+\tau$ which is our goal in this section.

Following Forrester,  consider $E_N(x)= \Prob\left\{ x_{max} (X) \le x \right\}$. This is the probability for matrix $X$ to have no eigenvalues to the right of the vertical line $\Re z=x$. Therefore,
 \begin{equation}\label{Probmax1}
E_N(x)=\Big\langle \prod_{j=1}^N \mathit{\Theta}(x-x_j)\Big\rangle_X \, ,
\end{equation}
where $x_j$ are the real parts of the eigenvalues of $X$, $x_j=\Re z_j$, and $\Theta (x)$ is the step function, Eq.~\ref{Heav}. 
By making use of the identity  $\mathit{\Theta}(x)=1-\mathit{\Theta}(-x)$, one can expand the product on the right-hand side in Eq.~\ref{Probmax1} in powers of $\mathit{\Theta}$,
\begin{eqnarray*}
    \prod_{j=1}^N \mathit{\Theta}(x-x_j) = \prod_{i=1}^N \big(1-\mathit{\Theta}(x_i-x)\big) =
 1- \sum_{i=1}^N \mathit{\Theta}(x_i-x)
+ \sum_{i\ne j}^N \mathit{\Theta}(x_i-x)\mathit{\Theta}(x_j-x) - \ldots \,  .
\end{eqnarray*}
The $k$-th term in this expansion involves $k$-tuples of eigenvalues and, therefore, its average over the realizations of $X$ is expressed in terms of the $k$-point eigenvalue correlation function:
\begin{eqnarray}\label{kpoint}
 E_N(x) = 1-\int_{\mathbb{C}}   \mathit{\Theta}\left(x_1-x\right)\, R_1(z_1)d^2z_1 + \frac{1}{2!}\!\int_{\mathbb{C}} \int_{\mathbb{C}} \mathit{\Theta}(x_1-x) \mathit{\Theta}(x_2-x) R_2(z_1,z_2) \,  d^2z_1d^2z_2 + \ldots\, .
 \end{eqnarray}
We note that in the real elliptic ensemble the mean density of eigenvalues vanishes exponentially fast in the limit $N\gg 1$ outside the support of the limiting elliptic eigenvalue distribution. We will verify this fact later, see Propositions \ref{prop2_1} and \ref{prop2_2}. Consequently, for each $x>1+\tau$,
\[
\int_{\mathbb{C}}  \mathit{\Theta}\left(x_1-x\right)\, R_1(z_1)d^2z_1\ll 1\quad \quad (N\gg 1)\, . 
\]
Now, setting aside mathematical subtleties related to estimation of the next order correction terms in Eq.~\ref{factorisationR}, the factorization of the eigenvalue correlation functions implies that the terms containing higher order correlation functions in Eq.~\ref{kpoint} are significantly smaller than the term containing $R_1(z)$. Therefore, 
\[
E_N(x)\sim 1- \int_{\mathbb{C}}   \mathit{\Theta}\left(x_1-x\right)\, R_1(z_1)d^2z_1 = 1 - { N} \int_x^{\infty}\int_{-\infty}^{\infty}\left\langle \rho_N(x_1,y_1)\right\rangle_X \,dx_1 dy_1 \quad \quad (N\gg 1, \,\, x> 1+\tau)
\]
and, hence,  
\begin{equation}\label{Probmax2}
p_N(x)  \sim  {N}  \int_{-\infty}^{\infty}\left\langle \rho_N(x,y)\right\rangle_X dy \quad \quad (N\gg 1, \,\, x> 1+\tau). 
\end{equation}

\medskip

In the real elliptic ensemble, the probability for matrix to have at least one real eigenvalue is non-zero and the mean density of eigenvalues  is given by two contributions arising from the densities of complex (non-real) and real eigenvalues: 
\begin{equation}\label{decomp}
\left\langle \rho_N(x,y)\right\rangle_X=\langle \rho_N^{(c)}(x,y)\rangle_X+\delta(y)\langle \rho_N^{(r)}(x)\rangle_X \, .
\end{equation}
Correspondingly,  we define two functions:
\begin{equation}\label{ProbmaxComp}
p_N^{(c)}(x )= N\int_{-\infty}^{\infty}\langle \rho_N^{(c)}(x,y)\rangle_X dy 
\end{equation}
and
\begin{equation}\label{ProbmaxReal}
p_N^{(r)}(x )= N\langle \rho_N^{(r)}(x)\rangle_X\, .
\end{equation}
The function $p_N^{(r)}(x)$ is the mean density of real eigenvalues. By integrating $p_N^{(r)}(x)$ over an interval of the real line one obtains the expected number of real eigenvalues of $X$ in this interval. And $p_N^{(c)}(x)$ is the density of real parts of complex (non-real) eigenvalues. By integrating $p_N^{(c)}(x)$ over an interval of the real line one obtains the expected number of non-real eigenvalues whose projections on the real line fall into this interval.

Eq.~\ref{Probmax2} implies that 
\begin{equation}\label{Probmax3}
 p_N(x )= \frac{d}{dx} \Prob\{x_{max} \le x \} \sim  p_N^{(c)}(x)+ p_N^{(r)}(x)\quad \quad  (N\gg 1, \,\, x> 1+\tau),
 \end{equation}
and the problem of estimating the probability of large deviations $\Prob\{x_{max} > 1+\tau \}$ is reduced to the asymptotic evaluation of the density of real eigenvalues and the density of real parts of complex eigenvalues in the large deviation region $x>1+\tau$. Such asymptotic evaluation is possible due to the availability of closed form expressions for  $\langle \rho_N^{(c)}(x,y)\rangle_X$ and $\langle \rho_N^{(r)}(x)\rangle_X$  in terms of Hermite polynomials, see ref.~\cite{ForNag2008}. These can be used to obtain $p_N^{(c)}(x)$ and $p_N^{(r)}(x)$ in the limit $N\gg 1$ in the entire range of values of $x$: in the bulk (inside the elliptic domain of the limiting eigenvalue distribution ), in the transition regions around the far-left and far-right points of the elliptic domain, and in the tails of the eigenvalue distribution (outside the elliptic domain).  These results have independent interest and are summarized in Propositions \ref{prop2_1} and \ref{prop2_2} below.

In the bulk and in the transition region around the spectral edge the mean density of real eigenvalues $p_N^{(r)}(x)$ was obtained in ref.~\cite{ForNag2008} and in the large deviation regime $x>1+\tau$  this density  was obtained in ref.~\cite{FyoKhor2016}\footnote{Unfortunately, the corresponding expressions (24)-(25) presented in \cite{FyoKhor2016} contained several misprints, in particular the constant term in (\ref{ratereal}) was missing and the spurious factor $\sqrt{\tau}$ appeared under the last logarithm, though the correct expressions as presented in Eqs  \ref{preexpreal}--\ref{ratereal} here were used for actual calculations. The correct formulas Eqs \ref{preexpreal},\ref{ratereal} appeared in ref.~\cite{{YF2016}}.
}. For completeness we summarize these results in the proposition below.

\begin{proposition} \label{prop2_1} In the limit $N\gg 1$:
\begin{itemize}
\item[(a)] For every $|x|<1+\tau$,
\[
p_N^{(r)}(x ) \sim \sqrt{\frac{N}{2\pi (1-\tau^2)}}\, .
\]
\item[(b)] For every $x>1+\tau$,
\begin{equation}\label{39b}
p_N^{(r)}(x ) \sim Q^{(r)}_N(x) e^{-N\Psi^{(r)}(x)}\, ,
\end{equation}
where
 \begin{eqnarray}\label{preexpreal}
Q^{(r)}_N(x)&=&\sqrt{\frac{N}{2\pi(1+\tau)}\frac{1}{\sqrt{x^2-4\tau}\, (x+\sqrt{x^2-4\tau})}}, \\
\label{ratereal}
\Psi^{(r)}(x)&=&-\frac{1}{2}+\frac{x^2}{2(1+\tau)}-\frac{1}{8\tau}\left(x-\sqrt{x^2-4\tau}\right)^2 -
\ln{\frac{x+\sqrt{x^2-4\tau}}{2}}\, .
\end{eqnarray}
\item[(c)] For every fixed real $\delta$,
\begin{equation}\label{trans1}
p_N^{(r)}\Big( (1+\tau)\big(1+\frac{\delta}{\sqrt{N}}\big)\Big) \sim  \frac{1}{2}  \sqrt{\frac{N}{2\pi (1-\tau^2)}} \Big( 1- \erf (\delta_{\tau} \sqrt{2}) +\frac{1}{\sqrt{2}} e^{-\delta_{\tau}^2} \big(1+\erf(\delta_{\tau})\big)\Big)
\end{equation}
where $\erf (\delta)=\frac{2}{\sqrt{\pi}}\int_0^{\delta} e^{-t^2} dt $ is the error function and
\[
\delta_{\tau}=\delta\sqrt{\frac{1+\tau}{1-\tau}}\, .
\]
\end{itemize}
\end{proposition}

\medskip

One can verify that the asymptotic law of the density of real eigenvalues in the transition region interpolates between the flat density of real eigenvalues in the bulk and the exponentially small density in the tail of the distribution of real eigenvalues. Indeed, in the limit $\delta \ll -1$ the right-hand side in Eq.~\ref{trans1} is asymptotically equal to $\sqrt{{N}/{(2\pi (1-\tau^2))}}$, matching the expression for $p_N^{(r)}(x)$  in the bulk. In the opposite limit  $\delta \gg 1$, the right hand side of Eq.~\ref{trans1} is asymptotically equal to
$
\sqrt{{N}/{(4\pi (1-\tau^2))}}\, e^{-\delta_{\tau}^2}.
$
This is the same asymptotic law that one obtains on replacing $x$ in Eqs~\ref{preexpreal}--\ref{ratereal} by $(1+\tau)\big(1+\frac{\delta}{\sqrt{N}}\big)$ and expanding the right-hand side of Eq.~\ref{39b} in the limit $\delta\ll 1$.

\medskip

Now, we turn our attention to the density $\langle \rho_N^{(c)}(x,y)\rangle_X$ of complex (non-real) eigenvalues. To the best of our knowledge this density has not been analyzed yet in the large deviation regime $x> 1+\tau$. As we need it to evaluate the integrated version of this density, Eq.~(\ref{ProbmaxComp}), we provide a brief outline of such an analysis below.

\medskip

Let $\psi^{(\tau)}_k(z)=e^{-\frac{z^2}{2(1+\tau)}}h^{(\tau)}_k(z)$ where $h^{(\tau)}_k(z)$, $k=0,1,,2 \ldots, $ are the rescaled Hermite polynomials
\begin{eqnarray*}
h^{(\tau)}_{k}(z) = \frac{1}{\sqrt{\pi}}\int_{-\infty}^{\infty}e^{-t^2}
\left(z\pm it\sqrt{2\tau}\right)^k\,dt
 = \frac{(\pm i\sqrt{N})^k}{\sqrt{\pi}}\sqrt{\frac{N}{2\tau}}\int_{-\infty}^{\infty}e^{-\frac{N}{2\tau}(u\pm iz)^2}
u^k\,du \, ,
\end{eqnarray*}
and
\begin{eqnarray*}
S_{\tau}^{(c)}(x,y) =
 \frac{1}{2(1+\tau)\sqrt{2\pi}}\,
\sum_{j=0}^{N-2}\, \frac{\psi^{(\tau)}_{j+1}(\overline{z})\psi^{(\tau)}_{j}(z)-\psi^{(\tau)}_{j}(\overline{z})\psi^{(\tau)}_{j+1}(z)}{j!}\, .
\end{eqnarray*}
Then, assuming  $N$ is even, (see Eq.~6.2 in ref.~\cite{ForNag2008})
\begin{eqnarray*}
\langle \rho_N^{(c)}(x,y)\rangle = 2i \sign (y) \erfc\!\left(\!\!\sqrt{\frac{2}{1-\tau^2}}|y|\sqrt{N}\!\right)S_{\tau}^{(c)}(x,y).
\end{eqnarray*}
By manipulating the integral representation for $h^{(\tau)}_{k}(z)$, one obtains 
 \begin{equation}\label{denscomp2}
S_{\tau}^{(c)}(x,y)= i\left(\frac{N}{2\pi}\right)^{\!3/2}\!\!\!\frac{1}{\sqrt{2}\, \tau(1+\tau)}\,\,I_N^{(c)}(x,y)
\end{equation}
with
\begin{eqnarray}\label{denscomp3}
I_N^{(c)}(x,y)= \int_{-\infty}^{+\infty} dp\, p \int_{-\infty}^{+\infty}dq \, e^{-N(A_y(p)+B_x(q))}\frac{\Gamma\!\left(\!N-1,\frac{N}{2}(p^2-q^2)\right)}{ \Gamma (N-1)}
\end{eqnarray}
where
\[
A_y(p)=\frac{1-\tau}{2\tau}p^2+\frac{p\sqrt{2}}{\tau}y , \,\,\, B_x(q)=\frac{1+\tau}{2\tau}q^2+\frac{iq\sqrt{2}}{\tau}x\, ,
\]
and
$\Gamma\left(N,a\right)$ is the incomplete Gamma function,
\begin{equation}\label{phiphi}
\Gamma (N,a)=\int_a^{\infty}\!\!e^{-t}\,t^{N-1}\,dt=\Gamma (N)\,  e^{-a}\sum_{k=0}^{N-1} \frac{a^k}{k!}\,.
\end{equation}
This function is well known to have the following limiting behaviour:
\begin{equation}\label{phiphilim}
\lim_{N\to\infty}\frac{\Gamma\left(N-1,Na\right)}{\Gamma (N-1)}=
\begin{cases}
1, &  \mbox{if   } 0\le a<1, \\
\frac{1}{2}, &  \mbox{if   } a=1, \\
0, & \mbox{if   }  a>1\, ,
\end{cases}
\end{equation}
and, moreover, in the transition region around $a=1$, 
\begin{equation}\label{phiphilim1}
\lim_{N\to\infty}\!\frac{\Gamma(N-1,N(1+\alpha N^{-1/2}))}{\Gamma (N-1)}=\frac{1}{\sqrt{2\pi}}\int_{\alpha}^{\infty}e^{-\frac{v^2}{2}}\,dv\, .
\end{equation}
In particular, evaluating the integrals in Eq.~\ref{denscomp3} by the saddle-point method, taking Eq.~\ref{phiphilim}
into account, and using the asymptotic relation
\begin{equation}\label{err}
 \mbox{erfc}\left(\sqrt{\frac{2}{1-\tau^2}}|y|\sqrt{N}\right)\sim \sqrt{\frac{1-\tau^2}{2\pi N}}\frac{1}{|y|}e^{-\frac{2}{1-\tau^2}Ny^2} \quad \quad (N\gg 1)
 \end{equation}
 one immediately reproduces  the elliptic law of Eq.~\ref{dmu_eq} for the limiting density $\lim_{N\to\infty}\langle \rho_N^{(c)}(x,y)\rangle$. Finer details of the distribution of the non-real eigenvalues of $X$,  like the density profile in the transition region around the elliptic boundary of the eigenvalue distribution can be obtained using Eq.~\ref{phiphilim1}.

Eqs~\ref{denscomp2}--\ref{denscomp3} come in handy for asymptotic analysis of the integrated density of complex eigenvalues $p_{N}^{(c)}(x)$, Eq.~\ref{ProbmaxComp}.

\begin{proposition} \label{prop2_2}In the limit $N\gg 1$:
\begin{itemize}
\item[(a)] For every $|x|<1+\tau$,
 \begin{equation}\label{bulklimit}
p_{N}^{(c)}(x) \sim \frac{2N}{\pi 1+\tau}\, \sqrt{1-\frac{x^2}{1+\tau^2}} \, .
\end{equation}
\item[(b)] For every $x>1+\tau$,
\begin{equation}\label{denscompintegrRight}
p_N^{(r)}(x ) \sim Q^{(c)}_N(x) e^{-N\Psi^{(c)}(x)}\, ,
\end{equation}
where  $\Psi^{(c)}(x)=2\Psi^{(r)}(x)$ and
\begin{equation*}
 Q^{(c)}_N(x)=
 \sqrt{\frac{N}{2(1+\tau)}}\,
 \frac{b^2(x)}{\pi \left(1-b(x)\right)^{3/2}\left(1-\tau b(x)\right)^{1/2}}\, , \quad \quad b(x)=\left(\frac{x-\sqrt{x^2-4\tau}}{2\tau}\right)^2\, .
\end{equation*}

\item[(c)] For every fixed real $\delta$,
\begin{equation}\label{part_c}
p_N^{(c)}\Big( 1+\tau\big(1+\frac{\delta}{\sqrt{N}}\big)\Big) \sim  \frac{2^{1/2}N^{3/4}}{\pi^{3/2}}\frac{1}{\sqrt{1-\tau^2}}\int_0^{\infty} e^{-\frac{1}{2}\frac{1+\tau}{1-\tau}(q+2\delta)^2} \sqrt{q}\,dq\, .
\end{equation}
\end{itemize}
\end{proposition}

\emph{Proof}.
The integration over $y$ in Eq.~\ref{ProbmaxComp} can be performed by making use of the asymptotic relation Eq.~\ref{err}. This  yields
\begin{eqnarray} \label{denscompintegr}
p_{N}^{(c)}(x) \sim \frac{N^{3/2}e^{N\frac{x^2}{\tau(1+\tau)}}}
{\pi^{3/2}\sqrt{\tau(1+\tau)}}
\int_{-\infty}^{\infty}dq\, e^{-N\left(q^2\frac{1+\tau}{2\tau}+i\frac{\sqrt{2}xq}{\tau}\right)}
\frac{1}{\Gamma (N-1)} \int_{0}^{\infty} dp\,  \Gamma\left(N-1,\frac{N}{2}\left(p^2-q^2\right)\right)\, .
\end{eqnarray}
On performing integration by parts in the $p$-integral, one transforms the right-hand side into a form suitable for saddle point analysis,
 \begin{equation}\label{denscompintegr1}
p_{N}^{(c)}(x) \sim
\frac{
{2N^{3/2} \left( \frac{N}{2} \right)^{N-1}} e^{ N\frac{x^2}{\tau(1+\tau)} }}{\pi^{3/2}\sqrt{\tau(1+\tau)}(N-2)!}
\int_{-\infty}^{\infty}dq \int_{0}^{\infty} \frac{dp\, p^2}{(p^2-q^2)^2}  \,
e^{-N\Big(\frac{q^2}{2\tau}+i\frac{\sqrt{2}xq}{\tau}\Big) -{N}\left(\frac{p^2}{2}-\ln{\frac{p^2-q^2}{2}}\right)}\, .
\end{equation}
Expectedly, the asymptotic behavior of the integrated density $p_{N}^{(c)}(x)$ is controlled by the ratio $\frac{|x|}{1+\tau}$. If $|x|<1+\tau$ then
the integrals in $p$ and $q$ in Eq.~\ref{denscompintegr1} 
are dominated by the saddle point at
\begin{equation*}
q=q_*=-i\frac{\sqrt{2}x}{1+\tau}, \quad
p=p_*=\sqrt{2\left[1- \frac{x^2}{(1+\tau)^2}\right]}\, ,
\end{equation*}
and the saddle-point analysis yields Eq.~\ref{bulklimit}.
Not surprisingly, this semicircular density can also be obtained by integrating the limiting elliptic eigenvalue distribution, see Eq.~\ref{dmu_eq}, over $y$, the imaginary part of eigenvalues.

\medskip

If $x>1+\tau$ then the integral in $p$ in Eq.~(\ref{denscompintegr1}) is dominated by the neighborhood of $p=0$, the lower boundary of the interval of integration, whereas the integral in $q$ is dominated by the saddle-point at $q=q_*=-\frac{i}{\sqrt{2}}\left(x+\sqrt{x^2-4\tau}\right)$. Applying the saddle-point method, one obtains, after a lengthy but straightforward computation Eq.~\ref{denscompintegrRight}.

\medskip

The asymptotic relation  Eq.~\ref{part_c} can be obtained from Eq.~\ref{denscompintegr}.  On changing variables of integration in the latter to
\[
u=N^{1/2}\left(\sqrt{\frac{1+\tau}{\tau}}q+i\frac{x\sqrt{2}}{\sqrt{\tau(1+\tau)}}\right), \quad w=pN^{1/4}\, ,
\]
one arrives at
\begin{eqnarray*}
p_{N}^{(c)}(x)\sim \frac{\sqrt{2}N^{3/4}}{\pi(1+\tau)\Gamma (N-1)}
\int_{-\infty}^{\infty}\frac{du}{\sqrt{2\pi}}\,  e^{-\frac{u^2}{2}}
\int_{0}^{\infty} \!\!\! dw \, \Gamma\left(N-1,N\left[\frac{x^2}{(1+\tau)^2}+\frac{\beta}{\sqrt{N}}+O(N^{-1})\right]\right),
\end{eqnarray*}
where 
$\beta=\frac{w^2}{2}+i\,u\,x\sqrt{\frac{2\tau}{(1+\tau)^3}}$.
Now, one can substitute here $\frac{x}{1+\tau}=1+\frac{\delta}{\sqrt{N}}, \, \frac{x^2}{(1+\tau)^2}=1+\frac{2\delta}{\sqrt{N}}+O(N^{-1})$ and employ Eq.~\ref{phiphilim1} with $\alpha=\frac{w^2}{2}+2\delta+iu\sqrt{\frac{2\tau}{1+\tau}}$ for extracting the leading asymptotic term of the incomplete gamma function in the limit $N\gg 1$.  Further, by making use of integration by parts in  $w$, one can evaluate the integral in $u$ in closed form to arrive, after simple manipulations, at Eq.~\ref{part_c}. \hfill $\Box$

\medskip

We would like to note that when $\delta$ runs from $-\infty$ to $+\infty$ the function 
\[
\tilde p_{N}^{(c)}(\delta)=p_{N}^{(c)}\left(1+\tau\Big(1+\frac{\delta}{\sqrt{N}}\Big)\right) 
\]
provides a smooth crossover from the semicircular density in 'bulk', Eq.~\ref{bulklimit}, to the exponential decay in the right tail of the distribution of real line projections of complex eigenvalues outside the limiting ellipse, Eq.~\ref{denscompintegrRight}. Indeed, by rescaling $q\to |\delta| q$ one can easily find the asymptotics of $\tilde p_{N}^{(c)}(\delta)$ in the limits  $\delta\to \pm \infty$:
\begin{eqnarray}\label{denscompintegredgepositive}
\hspace*{-4ex} \tilde p_{N}^{(c)}(\delta) &\sim& \frac{N^{3/4}(1-\tau)}{4\pi(1+\tau)^2}\delta^{-3/2} e^{-2\frac{1+\tau}{1-\tau}\delta^2}, \quad \delta\to +\infty,  \\
\label{denscompintegredgenegative}
\hspace*{-4ex} \tilde p_{N}^{(c)}(\delta) &\sim & \frac{2^{3/2}N^{3/4}}{\pi(1+\tau)}|\delta|^{1/2}\, , \hspace{13ex}  \quad\delta\to -\infty\, .
\end{eqnarray}
It is easy to check that (\ref{denscompintegredgenegative}) perfectly matches the semicircular density (\ref{bulklimit}). To see this,
substitute $x/(1+\tau)=1-\frac{|\delta|}{\sqrt{N}}$ in (\ref{bulklimit}) and expand. Similar, but a lengthier calculation shows that (\ref{denscompintegredgepositive}) perfectly matches both exponential and pre-exponential terms in Eq.~(\ref{denscompintegrRight})
on replacing $x/(1+\tau)=1+\frac{\delta}{\sqrt{N}}$ and expanding.

\medskip

With Eqs~\ref{39b} and  \ref{denscompintegrRight} in hand, the probability density function $p_N(x)$ of $x_{max}$  now follows from Eq.~\ref{Probmax3}. Since the rate of decay for the density of real parts of complex eigenvalues is twice the one for the density of real eigenvalues, the former gives no contribution in the leading order and 
\begin{equation}\label{LDP_B}
p_N(x) \sim Q^{(r)}_N(x) e^{-N\Psi^{(r)}(x)} \quad\quad  ( N\gg 1, \, x> 1+\tau)
\end{equation}
where the pre-exponential factor $Q^{(r)}_N(x) $ and the large deviation rate function $\Psi^{(r)}(x)$ are given in Eqs~ \ref{preexpreal}--\ref{ratereal}. Since $\Psi^{(r)}(x)$ is an increasing function of $x$, the integral in Eq.~\ref{bbb} will be dominated by the neighborhood of the lower boundary of the interval of integration, and we obtain the desired large deviation principle for $x_{max}$: 
\begin{equation}\label{LDP_B}
\tcboxmath{
 \Prob\{  x_{max} > x\}=e^{-N\Psi^{(r)}(x) +o(N)} \quad\quad   ( N\gg 1, \, x> 1+\tau).
}
\end{equation}

\section{Large  Deviation Principle for the eigenvalue counting measure in the real elliptic ensemble}
\label{subsection_C}

In this section we find the probability of large deviations of the eigenvalue counting measure $\mu_N$, Eq.~\ref{dmu_N}, from the Elliptic Law $\mu_{eq}$, Eq.~\ref{dmu_eq}. Although the large deviation principle for the \emph{complex} elliptic ensemble was established in \cite{PetzHiai98}, it is not immediately obvious to us how to extend this result to the real elliptic ensemble. This is because
the finite-$N$ structure of the eigenvalue distribution for real matrices differs from that in the complex case (real matrices have a non-zero density of real eigenvalues). Instead, we extend the large deviation principle for the real Ginibre ensemble ($\tau=0$) of ref.~\cite{BZ1998} to the elliptic case ($0<\tau<1$).

The real Ginibre ensemble is defined on the space of real $N\times N$ matrices by the probability measure with the density
\begin{equation*}
  {\cal P}_{\alpha}(Z)=\frac{1}{{\cal Z}_{N}(\alpha)} e^{-\alpha \frac{ N}{2}\Tr Z\,Z^T} \quad \quad (\alpha>0)\, ,
\end{equation*}
where ${\cal Z}_{N}(\alpha)$ is the normalization constant. The matrix entries of $Z$ are  independent identically distributed real Gaussians with mean value zero and variance $\sqrt{\alpha N}>0$. Ben Arous and Zeitouni proved in ref.~\cite{BZ1998}  that the normalized eigenvalue counting measure $\mu_N$ associated with matrices $Z$ obeys a large deviation principle with speed $N^2$ and rate functional
 \begin{eqnarray*} 
{\cal J}_{\alpha} [\mu] = \frac{\alpha}{2} \int_{\mathbb{C}} |z|^2 \,  d\mu (z)
- \frac{1}{2}\int_{\mathbb{C}}\int_{\mathbb{C}}\ln |z-w|  \, d\mu (z) \,
d\mu (w) - \frac{3}{8}\alpha^2 \, .
\end{eqnarray*}
\begin{tcolorbox}
Loosely speaking, this means that if $B$ is a subset of ${\cal P}_{s}(\mathbb{C})$, the set of all probability measures on $\mathbb{C}$ which are symmetric with respect to the operation of reflection in the real axis, then
\begin{equation}\label{LDPginibre}
\Prob \{\mu_N \in   B\} \approx \exp\left\{-N^2\inf_{\mu\in B} \mathcal{J}_{\alpha}[\mu]\right\}\, ,
\end{equation}
where the symbol $\approx $ stands for asymptotic equality ignoring both the pre-exponential multiplicative terms and sub-leading additive terms in the exponential.
\end{tcolorbox}

The rate functional $ {\cal J}_{\alpha} [\mu] $ is strictly convex on ${\cal P}_{s}(\mathbb{C})$ and its unique global minimizer is the Circular Law, the uniform distribution on the disk of radius $\sqrt{\alpha}$. In this context, in order to understand the likely distribution of eigenvalues of $Z$ conditioned by a rare event, e.g., `` all eigenvalues lie to the left of the line $\Re z = x$, $x<\sqrt{\alpha}$'',  it will be enough to find the minimizer of ${\cal J}_{\alpha}$ on such an event.

The large deviation principle (\ref{LDPginibre}) can be extended to the elliptic case by means of the Laplace-Varadhan theorem from large deviations theory, see ref.~\cite{DemboZeit}, and, also, ref.~\cite{Touchette}  for a useful informal account. This theorem asserts that if a sequence of probability laws $P_N$ defined on the same sample space $\Omega$ obey a large deviation principle with speed  $\alpha_N>0$ and good rate function $0\le I(x)<\infty$, i.e.,
\begin{equation*}
P_N(B)\approx \exp{\left[-\alpha_N \inf_{\omega \in B} I [\omega ] \right]}, \quad B \subseteq \Omega,
\end{equation*}
then the sequence of probability laws 
 \begin{equation}\label{tilted1}
dQ_N (\omega)=\frac{1}{{\cal Z}_N}\,e^{-\alpha_N F(\omega)} dP_N(\omega), \quad
\end{equation}
obeys the large deviation principle with speed $\alpha_N$ and the rate functional $(F+I)-\inf_{\Omega}(F+I)$. 
The normalization constant in Eq.~\ref{tilted1},
\[
{\cal Z}_N=\int e^{-\alpha_N F(\omega)} dP_N (\omega)
\]
satisfies the relation
\begin{equation*}
\lim_{N\to \infty}\alpha_N^{-1}\log{{\cal Z}_N}=- \inf_{\omega \in \Omega}(F(\omega)+I(\omega))
\end{equation*}
which explains the appearance of $\inf_{\Omega}(F+I)$ in the rate function for $Q_N$. In simple cases like real-valued random variables, the assertion of the Laplace-Varadhan theorem follows from applying the Laplace method to the integral
\begin{equation}\label{tilted2}
Q_N(B)={{\cal Z}_N}^{-1}\int_B\,e^{-\alpha_N\,F(\omega)} dP_N(\omega)
\end{equation}
but the theorem holds true for much broader class of  sample spaces (e.g., for random measures).

Now, consider the real elliptic ensemble defined by the matrix distribution with two-parameter density (cf. Eq.~\ref{JPDGin})
 \begin{equation}\label{Zellip}
  {\cal P}_{\alpha,\kappa}(Z)=\frac{1}{{\cal Z}_{N}(\alpha,\kappa)} e^{-\frac{N}{2}  \Tr (\alpha Z\,Z^T-\kappa Z^2)}\, .
\end{equation}
If $\alpha=\frac{1}{1-\tau^2}$ and  $\kappa=\frac{\tau}{1-\tau^2}$ then ${\cal P}_{\alpha,\kappa}(Z)$ reduces to Eq.~\ref{JPDGin}. The right-hand side of Eq.~\ref{Zellip} is suggestive of an application of the Laplace-Varadhan theorem. Indeed, denote by $P_N^{(\alpha,\kappa)}$ the probability law on ${\cal P}_{s}(\mathbb{C})$, induced by the above matrix distribution (e.g., via approximation by atomic measures) and by $\langle\ldots \rangle_{\alpha,\kappa}$ the operation of averaging over $P_N^{(\alpha,\kappa)}$. In these notations our main object of interest, $P_N^{(\alpha,\kappa)}(B)=\Prob\{\mu \in B\}$,  can be written as
\[
P_N^{(\alpha,\kappa)}(B)= \frac{1}{{\cal Z}_{N}(\alpha,\kappa)} \int_B e^{-N^2 F[\mu]} dP_N^{(\alpha,0)} (\mu),
\]
where the functional $F$ is defined on atomic measures  by
\begin{equation}\label{F}
F[\mu_N]=-\frac{1}{2N}\, \kappa  \Tr Z^2=-\frac{1}{2}\, \kappa \int_{\mathbb{C}} \Re ( z^2)\,d\mu_N (z) \,
\end{equation}
and extended to the whole of  ${\cal P}_{s}(\mathbb{C})$ by continuity.  Alternatively, we can write $\Prob\{\mu_N \in B\}$ in terms of the normalized eigenvalue counting measure associated with the ensemble of Eq.~\ref{Zellip}:
\[
\Prob\{\mu_N \in B\}= \langle 1_{\mu_N\in B} \rangle_{\alpha,\kappa} \, ,
\]
where $1_{\mu_N\in B}$ is the indicator function of the event $ \mu_N \in B$, i.e., $1_{\mu_N\in B}=1$ if $ \mu_N \in B$ and $1_{\mu_N\in B}=0$ otherwise.
Note that
 \begin{equation*}
 \langle 1_{\mu_N\in B}\rangle_{\alpha,\kappa}
 = \frac{\left\langle e^{-N^2 F[\mu_N]}1_{\mu_N\in B} \right\rangle_{\alpha,0}}{\left\langle e^{-N^2 F[\mu_N]}\right\rangle_{\alpha, 0}}\, ,
 \end{equation*}
so  that the relation between the real elliptic and real circular ensembles is exactly of the form of the Laplace-Varadhan theorem, Eqs~\ref{tilted1} and \ref{tilted2}, and the large deviation principle for the two parameter real elliptic ensemble of Eq.~\ref{Zellip} follows with the rate functional:
 \begin{equation*} 
{\cal J}_{\alpha,\kappa}[\mu]={\cal J}_{\alpha}[\mu]-\frac{1}{2}\kappa\int_C \mbox{\small Re}\left( z^2\right)\,d\mu + const\, .
\end{equation*}
Setting here $\alpha=\frac{1}{1-\tau^2}$,  $\kappa=\frac{\tau}{1-\tau^2}$  one obtains the desired large deviation rate functional
for the real elliptic ensemble of Eq~\ref{JPDGin}:
\begin{eqnarray} \label{LDPel}
\tcboxmath{
{\cal J}_{\tau}[\mu]= \frac{1}{2}\int_{\mathbb{C}} \left( \frac{x^2}{1+\tau} +
\frac{y^2}{1-\tau} \right) d\mu (z)
- \frac{1}{2} \int_{\mathbb{C}}\int_{\mathbb{C}}\ln |z-w| \, d\mu (z)\,
d\mu (w)- \frac{3}{8}\, ,
}
\end{eqnarray}
where the value of the constant $C=3/8$ turns out to be independent of $\tau$ and can be established by comparing to the large deviation principle for the counting eigenvalue measure in the complex elliptic matrices which was established by
Petz and Hiai in ref.~\cite{PetzHiai98}.  Indeed, in the latter paper it was shown that the global unique minimizer for ${\cal J}_{\tau}[\mu]$
on the space of probability measures with no symmetry condition is given by the uniform distribution on the ellipse with half-axes $1\pm \tau$, in full agreement with Eq.~\ref{dmu_eq}.  Obviously, the symmetry constraint  in the case of real matrices does not change the minimizer.

\section{Logarithmic potential of the Elliptic Law}
\label{subsection_D}
In this section we obtain the logarithmic potential $ \mathit{\Phi} (x; \mu_{eq})$
of the elliptic distribution $d\mu_{eq}=\rho_{eq} (z) d^2z$ in closed form for every real $x$. To this end, we note that 
\begin{equation*}
\mathit{\Phi} (x; \mu_{eq})=\int_{\mathbb{C}} \ln{|x-z|}\,d\mu_{eq}(z)=\frac{1}{\pi (1-\tau^2)}\Re\! \int_D \ln (x-s-it)ds\,dt,
\end{equation*}
where $D$ is the elliptic domain $s^2/(1+\tau)^2+t^2/(1-\tau)^2\le 1$. 
On changing to the polar coordinates 
$s=(1+\tau)r\cos{\theta}, t=(1-\tau)r\sin{\theta}$ in $D$, one obtains
\begin{equation*}
\mathit{\Phi} (x; \mu_{eq})\!=\!\frac{1}{\pi}\Re \! \int_0^1\!\! r\,dr\!\! \int_0^{2\pi} \!\!\!  d\theta \, \ln{\big(x-(1+\tau)r\cos{\theta}-i(1-\tau)r\sin{\theta}\big)}\, .
  \end{equation*}
Differentiating over $x$ we correspondingly get
\begin{equation}\label{lp1}
\frac{d}{dx}\mathit{\Phi} (x; \mu_{eq})=\frac{1}{\pi}\Re \! \int_0^1\!\! r\,dr\,\, I(x;r,\tau) \,
\end{equation}
where we defined
 \begin{equation}\label{lp2}
I(x;r,\tau)=  \int_0^{2\pi} \!\!\!  \frac{d\theta}{x-(1+\tau)r\cos{\theta}-i(1-\tau)r\sin{\theta}\big)}
\end{equation}
To evaluate the latter integral assuming $x>0$ we rewrite it as a contour integral over the unimodular complex variable
$z=e^{i\theta}$, that is
\begin{equation}\label{lp3}
I(x;r,\tau)=  -i\oint_{|z|=1} \,\,  \frac{d z/z}{x-(1+\tau)r(z+z^{-1})/2-(1-\tau)r(z-z^{-1})/2}=
\frac{i}{r}\oint_{|z|=1} \,\,  \frac{d z}{z^2-\frac{x}{r}z+\tau}
\end{equation}
The denominator has zeroes at $z_{\pm}=\frac{1}{2}\left(\frac{x}{r}\pm \sqrt{\frac{x^2}{r^2}-4\tau}\right)$
which allows us to rewrite
\begin{equation}\label{lp4}
I(x;r,\tau)=  \frac{i}{r(z_{+}-z_{-})}\left(\oint_{|z|=1} \,\,  \frac{d z}{z-z_{+}}-\oint_{|z|=1} \,\,  \frac{d z}{z-z_{-}}\right)
\end{equation}
showing that to have a nonzero result one of the zeroes should be inside and the other outside of the unit circle.
The zeroes are either real or complex conjugate. In the latter case they are both simultaneously inside the circle
due to the relation $z_{+}z_{-}=\tau<1$. In the former case it is easy to check that the zeroes are both inside
the circle as long as $x<r(1+\tau)$ whereas for $x> r(1+\tau)$ we have $|z_{-}|<1$ but $|z_{+}|>1$. This is the only
parameter range when $I(x;r,\tau)$ has a non-vanishing value. Evaluating, we finally find:
\begin{equation}\label{lp5}
I(x;r,\tau)=  \left\{\begin{array}{cc}\frac{2\pi}{\sqrt{x^2-4\tau r^2}} & \quad \mbox{if} \,\,\,\,  x>r(1+\tau)
\\ 0  & \quad \mbox{if} \,\,\,\,  0<x<r(1+\tau)\end{array}
\right.
\end{equation}
Note the jump discontinuity at $x=r(1+\tau)$. Now coming back to equation (\ref{lp1}) we notice that taking $ x>1+\tau$ implies that for any $r<1$ we necessarily have $x>r(1+\tau)$, so that using (\ref{lp1}) we easily find
\begin{equation}\label{lp6}
\frac{d}{dx}\mathit{\Phi} (x>1+\tau; \mu_{eq})=\frac{1}{\pi}\Re \! \int_0^1\!\! r\,dr\,\,  \,\frac{2\pi}{\sqrt{x^2-4\tau r^2}}=\frac{x-\sqrt{x^2-4\tau}}{2\tau}
\end{equation}
On the other hand, when  taking $0< x<1+\tau$ we have instead that the integrand is nonzero only as long as $r<x/(1+\tau)$, implying in this range
\begin{equation}\label{lp7}
\frac{d}{dx}\mathit{\Phi} (0<x<1+\tau; \mu_{eq})=\frac{1}{\pi}\Re \! \int_0^{\frac{x}{1+\tau}}\!\! r\,dr\,\,  \,\frac{2\pi}{\sqrt{x^2-4\tau r^2}}=\frac{x}{1+\tau}
\end{equation}
Note the continuity at $x=(1+\tau)$. Finally, to restore the logarithmic potential $\mathit{\Phi} (x;\mu_{eq})$ from its derivative, we notice that the right-hand side in (\ref{lp6}) coincides with the derivative
 of the function
 \begin{equation}\label{Phi1outside}
\mathit{\Phi}_1(x)=\frac{1}{8\tau}\left(x-\sqrt{x^2-4\tau}\right)^2+\ln{\frac{x+\sqrt{x^2-4\tau}}{2}}\, .
\end{equation}
Since these two functions are equal at  $x=+\infty$, one concludes that $\mathit{\Phi} (x; \mu_{eq})= \mathit{\Phi}_1(x)$ for every $x\ge 1+\tau$. In particular, $\mathit{\Phi} (x=1+\tau; \mu_{eq})=\tau/2$. On the other hand, integrating (\ref{lp6}) and
assuming the continuity of the log-potential at $x=1+\tau$ immediately gives
\begin{equation}\label{Phi1inside}
\mathit{\Phi} (x; \mu_{eq})=\frac{x^2}{2(1+\tau)}-\frac{1}{2}, \quad 0\le x <1+\tau\, .
\end{equation}
Because of the symmetry of the elliptic law, $\mathit{\Phi} (x; \mu_{eq}) = \mathit{\Phi} (-x; \mu_{eq})$.  Thus:
\begin{eqnarray}\label{LinearStatAA}
\tcboxmath{
\mathit{\Phi} (x; \mu_{eq})= \int_{\mathbb{C}} \ln{|x-z|}\,d\mu_{eq}(z)=\begin{cases}
\frac{x^2}{2(1+\tau)}-\frac{1}{2}, & \mbox{if } |x| \le  1+\tau,\\[1ex]
\frac{1}{8\tau}\left(|x|-\sqrt{x^2-4\tau}\right)^2+\ln{\frac{|x|+\sqrt{x^2-4\tau}}{2}}, & \mbox{if } |x| \ge 1 +\tau.
\end{cases}
}
\end{eqnarray}

\section{Asymptotic Analysis of $\mathcal{D}_N(x)$ and $\mathcal{D}^{(\alpha)}_N(x)$ }
\label{subsection_E}

We start with $\mathcal{D}_N(x)$. Here, there are two cases to analyze: $x>1+\tau$ and $x<1+\tau$, where  $x=1+\tau$ is the right-most point of the elliptic domain of the limiting eigenvalue distribution of $X$.

\medskip

If $x>1+\tau$ then, as was established in Section  \ref{subsection_B},  the event that 
$x_{max}(X)> x$ is exponentially rare in the limit $N\gg 1$ and, for typical realizations of $X$, the constraint $x_{max}(X)< x$ in Eq.~\ref{b_xmax} is satisfied.
By making use of the equation $\mathit{\Theta} (x-x_{max}(X))=1-\mathit{\Theta} (x_{max}(X)-x)$, one can transform $\mathcal{D}_N(x)$ to a form which, in this case, is more convenient for the asymptotic analysis:
\begin{equation} \label{LinearStat1}
\mathcal{D}_N(x)= \big\langle e^{N\mathit{\Phi} (x; \mu_N)} \big\rangle_X\, -  \big\langle \mathit{\Theta} (x_{max}(X)-x)\, e^{N\mathit{\Phi} (x; \mu_N)} \big\rangle_X \, .
\end{equation}
Here $\mathit{\Phi} (x; \mu)$ is the log-potential of $\mu$, see Eqs~\ref{dmu_N}--\ref{Phi_N}.
    If $\delta \mathit{\Phi}   = \mathit{\Phi} (x; \mu_N) - \mathit{\Phi} (x; \mu_{eq})$ then the first term on the right-hand side is $\langle e^{N\mathit{\Phi} (x; \mu_N)} \rangle_X = e^{N\mathit{\Phi} (x; \mu_{eq})}  \langle e^{N\delta\mathit{\Phi}_N } \rangle_X$. Fix an $\varepsilon >0$ and write
\begin{equation}\label{c1}
 \big\langle e^{N\delta \mathit{\Phi }} \big\rangle_X = \int_{|\delta\mathit{\Phi}|\le \varepsilon} e^{N\delta \mathit{\Phi}  } dP(\mu_N)  + \int_{|\delta\mathit{\Phi}|>\varepsilon } e^{N\delta \mathit{\Phi}  } dP(\mu_N)\, .
 \end{equation}
 Then, the first term on the right-hand side in Eq.~\ref{c1} is at most  $e^{N\varepsilon} $ and the second term is $e^{ O(N)}\Prob \{ |\delta\mathit{\Phi} |>\varepsilon  \}$. By the large deviation principle for $\mu_N$ of Section \ref{subsection_C}, $\Prob \{ |\delta\mathit{\Phi} |>\varepsilon  \}=e^{-N^2 C_{\varepsilon}+ o(N^2)}$, with $C_{\varepsilon}>0$, so that the second term vanishes in the limit $N\gg 1$. As $\varepsilon$ can be taken arbitrary small, 
$\langle e^{N\delta \mathit{\Phi} } \rangle_X= e^{o(N)}$ and, therefore,
\begin{equation}\label{unconstrained}
\big\langle e^{N\mathit{\Phi} (x; \mu_N)} \big\rangle_X =  e^{N\mathit{\Phi} (x; \mu_{eq})+o(N)}\, .
\end{equation}
The second term on the right hand side in Eq.~\ref{LinearStat1} is
\[
 \big\langle \mathit{\Theta} (x_{max}(X)-x)\, e^{N\mathit{\Phi} (x; \mu_N)} \big\rangle_X = e^{N\mathit{\Phi} (x; \mu_{eq})}  \big\langle \mathit{\Theta} (x_{max}(X)-x)\, e^{N\delta\mathit{\Phi}_N }  \big\rangle_X\, .
\]
By the Cauchy inequality and Eq.~\ref{LDP_B},
\[
 \big\langle \mathit{\Theta} (x_{max}(X)-x)\, e^{N\delta \mathit{\Phi}  } \big\rangle_X^2\le  \Prob\{  x_{max} (X) > x\} \big\langle e^{2N\delta \mathit{\Phi} } \big\rangle_X = e^{-N\mathit{\Psi}^{(r)}(x) +o(N)} \big\langle e^{2N\delta \mathit{\Phi} } \big\rangle_X\, . 
\]
By employing the same argument as above, $\langle e^{2N\delta \mathit{\Phi} } \rangle_X= e^{o(N)}$. Hence
\begin{equation}\label{c2}
 \big\langle \mathit{\Theta} (x_{max}(X)-x)\, e^{N\mathit{\Phi} (x; \mu_N)} \big\rangle_X \le  e^{-\frac{1}{2}N\mathit{\Psi}^{(r)}(x)} e^{N\mathit{\Phi} (x; \mu_{eq})+o(N)} , \quad x> 1+\tau.
\end{equation}
On comparing Eqs~\ref{unconstrained} and \ref{c2}, one concludes that
\begin{equation}\label{DNA}
\tcboxmath{
\mathcal{D}_N(x) = e^{N\mathit{\Phi} (x; \mu_{eq}) +o(N)} \quad\quad  (x> 1+\tau).
}
\end{equation}

\medskip

If $x<1+\tau$ then, typically, $X$ will have a macroscopic number of eigenvalues located right of the vertical line $\Re z=x$ and only in very rare realizations of $X$ ($e^{-N^2} $-rare) the constraint $x_{max}<x$ is satisfied. Following the lore of the large deviation theory, in the limit $N\gg 1$, such realizations will have the same eigenvalue distribution $d\nu_x$ which is the minimizer of the large deviation rate functional, Eq.~ \ref{LDPel}, on the set $B_x^{(0)}$ of all symmetric with respect to complex conjugation probability distributions on the complex plane 
whose support lies left to the vertical line $\Re z=x$, i.e.
\begin{equation}\label{B_x}
B_x^{(0)}=\left\{\mu \in {\cal P}_s(\mathbb{C}):  \mu (\mathbb{H}_x) = 0 \right\}\, .
\end{equation}
Correspondingly, 
\begin{equation}\label{factorisation}
\tcboxmath{
\mathcal{D}_N(x)= e^{N\mathit{\Phi}(x; \nu_x)+o(N)} \Prob\{ x_{max } <x\} \quad\quad (x<1+\tau)\, .
}
\end{equation}
From the mathematical viewpoint, the factorization property above can be justified by making use of the large deviation principle for the empirical eigenvalue counting measure $\mu_N$. Indeed,
\begin{equation}\label{oN}
\frac{
\mathcal{D}_N(x)
}
{ e^{N\mathit{\Phi}(x; \nu_x)}  \Prob\{ x_{max } <x\}}
=
\frac{
\big\langle \mathit{\Theta} (x-x_{max}(X)) e^{ N\delta \mathit{\Phi}  }
\big\rangle_X
}
{ \Prob\{ x_{max } <x\} } \, ,
\end{equation}
where now
\[
\delta \mathit{\Phi}  = \mathit{\Phi}_N (x; \mu_N) - \mathit{\Phi}(x; \nu_x) \, .
\]
For every positive $\varepsilon$
\begin{eqnarray} \nonumber
\big\langle \mathit{\Theta} (x-x_{max}) e^{ N\delta \mathit{\Phi}  }
\big\rangle_X  &=&   \int_{|\delta\mathit{\Phi} |\le \varepsilon} \mathit{\Theta} (x-x_{max}) e^{N\delta \mathit{\Phi} } dP(\mu_N)  + \int_{|\delta\mathit{\Phi} |>\varepsilon }\mathit{\Theta} (x-x_{max})  e^{N\delta \mathit{\Phi}  } dP(\mu_N)
\\[1ex]
\label{00}
&\le & e^{N\varepsilon } \Prob\{ x_{max } <x\}  + e^{O(N)}  \Prob\{ x_{max } <x \text{ and } |\delta\mathit{\Phi} |>\varepsilon\}
\end{eqnarray}
It is apparent from the large deviation principle for  $\mu_N$ that
$
\Prob\{ x_{max } <x \text{ and } |\delta\mathit{\Phi} |>\varepsilon\} = e^{-N^2 C_{\varepsilon}+o(N)} \Prob\{ x_{max } <x \}
$
for some $C_{\varepsilon}>0$. This nullifies the $e^{O(N)}$ factor in the second term on the right-hand side in Eq.~\ref{00}. Since $\varepsilon$ can be taken arbitrary small, one concludes that the right-hand side in Eq.~\ref{oN} is $e^{o(N)}$ and, hence, Eq.~\ref{factorisation} holds.

\medskip

Now, consider $\mathcal{D}^{(\alpha)}_N(x)$,
\begin{equation*}
\mathcal{D}_N^{(\alpha)}\!(x)\!= \big\langle \mathit{\Theta} (\alpha- \mu_N(\mathbb{H}_x))\,  e^{N \mathit{\Phi} (x; \mu_N)}  \big\rangle_X=    \int_{{\cal P}_s(\mathbb{C})} \mathit{\Theta} (\alpha- \mu(\mathbb{H}_x)) e^{N \mathit{\Phi} (x; \mu)} dP_N(\mu) \, , 
\end{equation*}
where $dP_N(\mu)$ is the probability distribution on ${\cal P}_s(\mathbb{C})$ induced by the probability law for $\mu_N$ in the real elliptic ensemble and $\mathit{\Phi} (x; \mu)$ is the log-potential of $\mu$.

It is instructive to take a closer look at the probability
\[
\Prob \{ \mu_N(\mathbb{H}_x) \le \alpha \} =  \big\langle \mathit{\Theta} (\alpha\!-\! \mu_N(\mathbb{H}_x)) \big\rangle_X\, .
\]
By the large deviation principle for $\mu_N$,
\[
\Prob \{ \mu_N(\mathbb{H}_x) \le \alpha \} = \exp \Big[ {-N^2\!\!\! \inf_{\mu \in B_x^{(\alpha)} } J[\mu] +o(N^2)} \Big] \, ,
\]
where (cf Eq.~\ref{B_x})
\begin{equation}\label{B_x_alpha}
B_x^{(\alpha)}=\left\{\mu \in {\cal P}_s(\mathbb{C}):  \mu (\mathbb{H}_x) \le \alpha \right\}\, .
\end{equation}
Recall that the limiting eigenvalue distribution $d\mu_{eq}$ is the unique global minimizer of the rate functional $J_{\tau}$  and
$J_{\tau}[\mu_{eq}]=0$. Furthermore,
\[
 \mu_{eq} (\mathbb{H}_x)= \int_{x}^{+\infty} \!\!\!\!\!\!ds \int_{-\infty}^{+\infty}  \!\!\!\! dt \,\, \rho_{eq}(s, t) =
 \begin{cases}
 1,  & \text{for } x\le -(1+\tau) \, ,\\[0.5ex]
 \displaystyle{
 \frac{2}{\pi} \int_{\frac{x}{1+\tau }}^1 \sqrt{ 1 - t^2  } \, dt
 }  \, ,
 &  \text{for } |x|\le 1+\tau \, ,\\[0.5ex]
 0, & \text{for } x\ge 1+\tau\, .
 \end{cases}
\]
Therefore, if $x(\alpha)$ is the solution of the equation
\begin{equation}\label{x_alpha}
\alpha =   \frac{2}{\pi} \int_{\frac{x}{1+\tau }}^1 \sqrt{ 1 - t^2  } \, dt
\end{equation}
for $x$ in the interval $|x| \le 1+\tau$ then $ \mu_{eq} \in B_x^{(\alpha)}$ for every $x>x(\alpha)$  and $ \mu_{eq} \notin B_x^{(\alpha)}$) for every $x< x(\alpha)$.
Hence,
\[
\Prob \{ \mu_N(\mathbb{H}_x) \le \alpha \} = e^{-N^2 C_1 +o(N^2)} \quad \quad (x<x (\alpha))
\]
and
\[
1-\Prob \{ \mu_N(\mathbb{H}_x) \le  \alpha \} = \Prob \{ \mu_N(\mathbb{H}_x) > \alpha \} =  e^{-N^2 C_2 +o(N^2)} \quad\quad  (x>x(\alpha))
\]
with $C_1, C_2>0$.

It is now apparent that there are two cases to consider when analyzing $\mathcal{D}^{(\alpha)}_N(x)$ in the limit $N\gg 1$. If $x>x(\alpha)$ then the event that there are more than $\alpha N$ eigenvalues of $X$ on the right to the vertical line $\Re z= x$ is very rare in the limit $N\gg 1$ and, for typical realizations of $X$, the constraint $\mu_N(\mathbb{H}_x) < \alpha$ in Eq.~\ref{NstmainA_2} is satisfied. Repeating almost verbatim the argument leading to Eq.~\ref{unconstrained} one concludes that in this case
\begin{equation}\label{a3}
\mathcal{D}_N^{(\alpha)} (x) = e^{N\mathit{\Phi}(x;\mu_{eq}) +o(N)}  \quad\quad  (x>x(\alpha))\, .
\end{equation}
On the other hand, if $x>x(\alpha)$ then, typically, $X$ will have more than $\alpha N$ eigenvalues on the right to the vertical line $\Re z= x$ and only in very rare realizations of $X$ the constraint $\mu_N(\mathbb{H}_x) < \alpha$ is satisfied.
Repeating almost verbatim the argument leading to Eq.~\ref{factorisation} one concludes that
\begin{equation}\label{factorisation_alpha}
\mathcal{D}_N^{(\alpha)}(x)= \exp\left[{N\mathit{\Phi}\big(x; \nu_x^{(\alpha)}\big)+o(N)}\right]\,  \Prob\{ \mu_N(\mathbb{H}_x) < \alpha\} \quad\quad  (x<x(\alpha))\, ,
\end{equation}
where $\nu_x^{(\alpha)}$ is the minimizer of the large deviation rate functional $J_{\tau}$ on the set $B_x{(\alpha)}$.

\medskip

To summarize findings in this Section: with the convention that $\mathcal{D}_N^{(0)} (x) = \mathcal{D}_N(x)$, for fixed $x$ and $\alpha$ in the limit $N\gg 1$,
\begin{equation} \label{E:LDoutcome}
\tcboxmath{
\mathcal{D}_N^{(\alpha)}(x) =  \left\{\begin{array}{lc} \exp \left[ {N\mathit{\Phi} (x; \mu_{eq}) +o(N) }\right]\, , &  \text{if }  x> x(\alpha), \\[1.5ex]
\exp\left[ {N\mathit{\Phi} (x; \nu_x^{(\alpha)}) +o(N) } \right] \, \Prob\{ \mu_N(\mathbb{H}_x) < \alpha\} , & \text{if } x< x(\alpha)\, .
\end{array}\right.
}
\end{equation}
Here $x(\alpha)$ is the solution Eq.~\ref{x_alpha} (note that $x(0)=1+\tau$), $\mathit{\Phi} (x; \mu_{eq})$ is the log-potential of the limiting elliptic eigenvalue distribution in the real elliptic ensemble, see Section \ref{subsection_D}, and $\mathit{\Phi} (x; \nu_x^{(\alpha)})$ is the log-potential of the limiting conditional eigenvalue distribution $d\nu_x^{(\alpha)}$ in the real elliptic ensemble, conditional on the event that $ x_{\alpha N+1} < x $, where $x_1, \ldots, x_N$ are ordered real parts of the eigenvalues, from the largest $x_1$  to the smallest $x_N$. The measure $\nu_x^{(\alpha)}$ is the
minimizer\footnote{Finding  such a minimizer in a closed form is a highly nontrivial exercise in potential theory, which for the present case is only solved for the special case of purely gradient flow $\tau=1$ and $\alpha=0$ \cite{DM_SP}, and partly characterized for $\tau=0$, $\alpha=0$   in ref.~\cite{ASZ2014}.
Fortunately, for our present purposes the exact form of the minimizer $\nu_x^{(\alpha)}$ is not needed.}
of the large deviation rate functional $J_{\tau}$ on the set $B_x^{(\alpha)}$, see Eq.~\ref{LDPel} and \ref{B_x_alpha}. Introducing the notation
\begin{equation}\label{ktau}
K_{\tau}^{(\alpha)}(x)
={\cal J}_{\tau}\big[\nu_x^{(\alpha)}\big] 
\end{equation}
so that
\begin{equation}\label{LDP3}
\tcboxmath{
\Prob \{ x_{\alpha N+1} <x\} =  \Prob\{ \mu_N(\mathbb{H}_x) < \alpha\} = e^{- N^2K_{\tau}^{(\alpha)}(x) + o(N^2)} \quad \quad (x<x(\alpha))\, .
}
\end{equation}
We note for future reference that $K_{\tau}^{(\alpha)}(x)$ is monotone decreasing function of $x$ on the interval $(-\infty, 1+\tau]$ which vanishes at $x=1+\tau$, $K_{\tau}^{(\alpha)}(1+\tau)=0$.

\section{Average number of stable equilibria} \label{subsection_F}

In this section we calculate the average number of the stable equilibria, $\langle{\cal N}_{st}\rangle$, in the limit $N\to\infty$ in the parameter range $0< m<1$ and $0<\tau<1$.

\medskip

It is instructive to evaluate first the average total number of all equilibria, $\langle{\cal N}_{eq}\rangle$. As was shown in ref.~\cite{FyoKhor2016},
\[
\langle{\cal N}_{eq}\rangle = \frac{1}{m^N} \int_{-\infty}^{\infty} \!   \big\langle |\det (X-xI)|   \big\rangle_X  \,  e^{-\frac{N(x-m)^2}{2\tau}} \frac{dx}{\sqrt{2\pi\tau/N}} \, .
\]
The equality here holds to leading order in $N$. Recalling Eq.~\ref{unconstrained},
\begin{equation}\label{Sigma_eq}
\langle{\cal N}_{eq}\rangle =  \int_{-\infty}^{\infty} \!   e^{N {\Sigma}_{eq} (x) +o(N)} \, dx\, , \quad \quad \text{where}\quad {\Sigma}_{eq}(x)= {\Phi} (x; \mu_{eq}) -\frac{(x-m)^2}{2\tau}-\ln m \, .
\end{equation}
This integral can be evaluated by the Laplace method. In the limit $N\gg 1$ this integral is dominated by the neighborhood of the global maximum of the function ${\Sigma}_{eq}(x)$ which can easily be determined by making use of Eqs.~\ref{lp6}  -- \ref{LinearStatAA} in Section~\ref{subsection_D}. It follows from Eq.~\ref{lp6} that for every $x\ge 1+\tau$
\[
{\Sigma}^{\, \prime}_{eq}(x)=\frac{2m - x-\sqrt{x^2-4\tau}}{2\tau}  \le \frac{m-1}{\tau}<0 \quad (0<m<1).
\]
Hence, the function ${\Sigma}_{eq}(x)$ is monotone decreasing on this interval. 
Similarly, since ${\Phi} (x; \mu_{eq}) $ is an even function of $x$,
\[
{\Sigma}^{\, \prime}_{eq}(x)=\frac{2m +|x|+\sqrt{x^2-4\tau}}{2\tau}  \ge \frac{m+1}{\tau} > 0\, 
\]
for every $x\le -(1+ \tau)$. Hence ${\Sigma}_{eq}(x)$ is monotone increasing on this interval, and
\begin{equation}\label{Sigma_eq2}
{\Sigma}_{eq}( -(1+ \tau)) >{\Sigma}_{eq}(x) \quad  \text{ for every } x<  -(1+ \tau)\, .
\end{equation}
\begin{equation}\label{Sigma_eq1}
{\Sigma}_{eq}(1+\tau) >{\Sigma}_{eq}(x) \quad  \text{ for every } x> 1+\tau\, .
\end{equation}
\begin{figure}[t!]
\centering
\includegraphics[width=.5\linewidth]{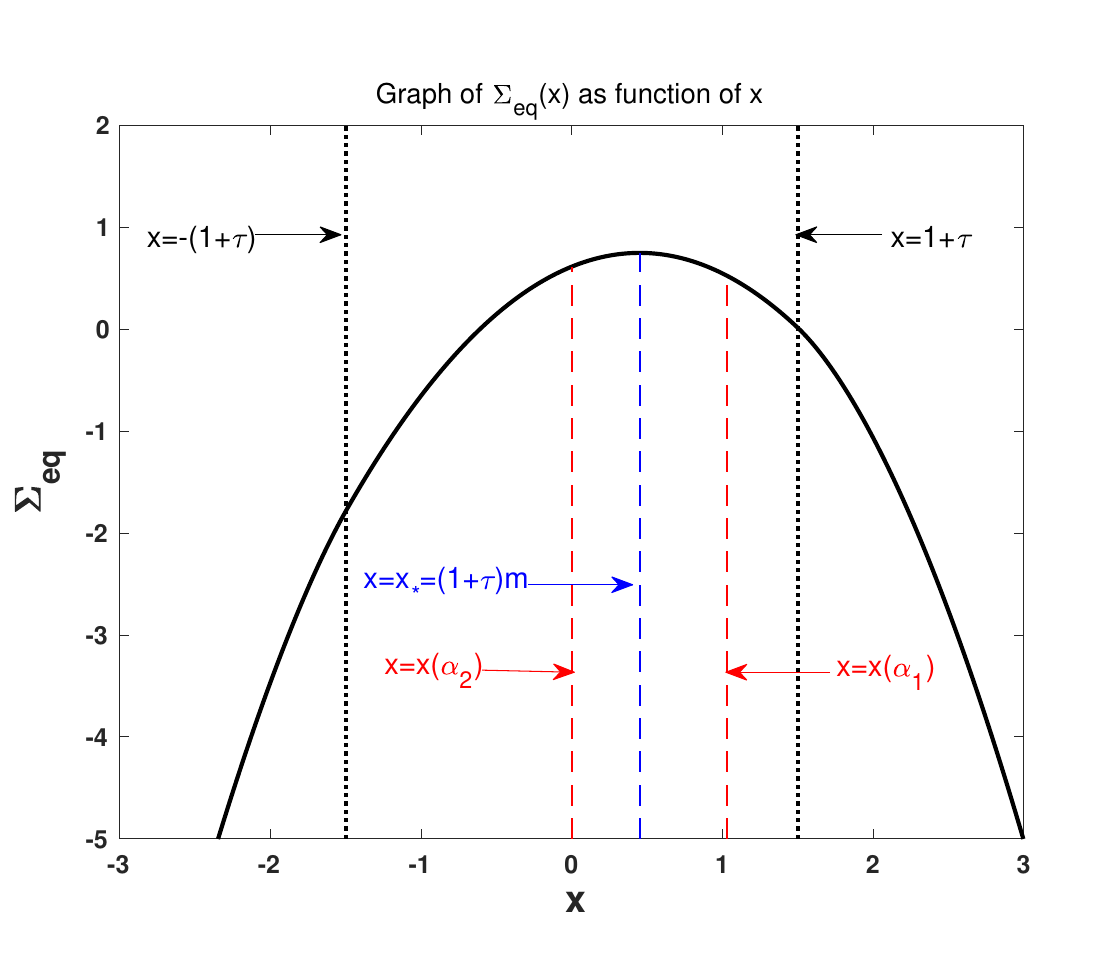}
\vspace*{-3ex}
\caption{The graph of $\mathit{\Sigma}_{eq} (x) $ as function of $x$ for parameter values $m=0.3$ and $\tau=0.5$. The two dotted lines show the boundaries of the elliptic eigenvalue distribution along the real line at $x=\pm (1+\tau)$. The blue dashed line shows the global maximum of  $\mathit{\Sigma}_{eq} (x) $  at $x=x_*=(1+\tau)m$. The two red dashed lines show the value of $\mathit{\Sigma}_{eq} $ at $x=x(\alpha)$, see Eq.~\ref{x_alpha}, for two values of $\alpha$, $\alpha_1=0.1$ and $\alpha_2=0.5$}
\label{fig:1}
\end{figure}
On the interval $|x|\le 1+\tau$, the function ${\Sigma}_{eq} (x)$ is quadratic, attaining its maximal value at the point $x=x_*=(1+\tau)m$ inside this interval. In view of Eqs.~\ref{Sigma_eq1}--\ref{Sigma_eq2} and the continuity of ${\Sigma}_{eq}(x)$, this maximal value is the unique global maximum of ${\Sigma}_{eq}(x)$,  see Fig.~\ref{fig:1},
\[
 \max_{x\in \mathbb{R}} {\Sigma}_{eq}(x) = {\Sigma}_{eq}((1+\tau)m) = \frac{m^2-1}{2} -\ln m \, .
\]
Hence, by the Laplace method,
\[
\lim_{N\to\infty}\frac{1}{N} \ln \langle{\cal N}_{eq}\rangle = \mathit{\Sigma}_{eq},
\]
where
\begin{equation}\label{Sigma_eq}
\mathit{\Sigma}_{eq} = \frac{m^2-1}{2} -\ln m  \quad (0<m<1)
\end{equation}
in agreement with ref.~\cite{FyoKhor2016} where this result was obtained by a different method.
\medskip

Now, we  turn our attention to $\langle{\cal N}_{st}\rangle$. The starting point of our analysis is Eq.~\ref{A:Nstmain} which holds to leading term in $N$. Guided by Eq.~\ref{E:LDoutcome}, we split the interval of integration in Eq.~\ref{A:Nstmain} in two domains $x> 1+\tau$ and $x< 1+\tau$.  Correspondingly,
\begin{equation}\label{Nstass}
\langle{\cal N}_{st}\rangle= I_{eq}{(N)}  +I_{st}{(N)} \, .
\end{equation}
Here
\begin{align}\label{NstassA}
I_{eq}{(N)}  &= \int_{1+\tau}^{\infty} e^{N {\Sigma}_{eq}(x) +o(N)} \, dx \, ,  & 
\\[1ex]
\label{NstassB}
I_{st}{(N)} &= \int_{-\infty}^{1+\tau}  e^{N{{\Sigma}_{st}}(x)+o(N)} \Prob\{ x_{max } <x \}
\, dx , & \text{where}\quad {\Sigma}_{st}(x)=\mathit{\Phi} (x;\nu_x) -\frac{(x-m)^2}{2\tau} -\ln m \, .
\end{align}
Both integrals can be evaluated in the limit $N\gg 1$ by the Laplace method.

\medskip

Let us first consider $I_{eq}{(N)}$. In the limit $N\gg 1$ this integral is
dominated by the neighborhood of the global maximum of the function ${\Sigma}_{eq}(x)$ on the interval $x\ge 1+\tau$. We have established above that ${\Sigma}_{eq}(x)$ is monotone decreasing on the interval $x\ge 1+\tau$. Hence,
\begin{equation}\label{NstassAfin}
\lim_{N\to\infty}\frac{1}{N} \ln I_{eq}{(N)}  = {\Sigma}_{eq}(1+\tau)  \quad \quad (0<m<1)\, .
\end{equation}

\medskip

Turning our attention to $I_{st}{(N)} $, the integral in Eq.~\ref{NstassB} requires a more careful approach.  Away from the upper limit of integration at $x=1+\tau$ (i.e., for every fixed $x<1+\tau$), the factor $\Prob \{ x_{max } <x\}$ in the integrand prevails due to its dominant scaling with $N$, see Eq.~\ref{LDP3}, i.e., the other factor of the integrand can be ignored.
However, as $\Prob \{ x_{max } <x\}$  is a monotonically increasing function of $x$, the integral in Eq.~\ref{NstassB} is dominated by the immediate neighborhood of its upper limit of integration at $x=1+\tau$ where the large deviation rate function $K_{\tau}(x) \equiv K_{\tau}^{(0)}(x) $ is actually vanishing, $K_{\tau}(1+\tau)=0$.  Hence, next-to-leading order corrections in Eq.~\ref{LDP3} cannot be ignored when evaluating $I_{st}{(N)}$. In other words,  for our goal of evaluating  the integral in Eq.~\ref{NstassB} the precision of Eq.~\ref{LDP3}  is not sufficient. What is actually needed is a sharper large deviation principle which includes the next sub-leading term in the exponential.

We conjecture that this term is of order $N$:
\begin{equation}\label{LDP4}
\tcboxmath{
\text{Conjecture: }\quad \Prob \{ x_{max} <x \} = e^{-N^2K_{\tau}(x)-NT_{\tau}(x)+o(N)}\,   \quad \quad (x<1+\tau).
}
\end{equation}
 Our conjecture is based on a similar sharper large deviation principle for the largest eigenvalue of Gaussian Hermitian and real symmetric matrices which was obtained in the framework of a powerful, albeit heuristic version of the Large Deviation Theory for random matrices  known as the 'Coulomb gas' method,
see calculations in, e.g.,  ref.~\cite{Borot_2011} and, closer to our context, in Appendix C of ref.~\cite{FyoWi07}.
 Although similar heuristic justifications for the validity of Eq.~\ref{LDP4} can be provided for our case as well,
a rigorous verification of such sharp large deviation principle and the problem of explicitly characterizing the function
 $T_{\tau}(x)$ remains a challenging task and is left for future research. It is clear, however,
 on general grounds that when approaching the boundary of the domain of the limiting eigenvalue distribution both functions $K_{\tau}(x)$ and $T_{\tau}(x)$
 must be vanishing. We shall assume\footnote{A simple scaling argument can be employed to conjecture the values of $p$ and $q$. to this end, note that the width of the transition region at the boundary of the elliptic eigenvalue distribution is proportional to $1/\sqrt{N}$, see Section \ref{subsection_B}. 
One would expect then that the distribution of $x_{max}$ in the transition region to the left of $1+\tau$ has a nontrivial finite shape. Correspondingly, setting $x=1+\tau -u/\sqrt{N}$, $u>0$, in Eq.~\ref{LDP4} and using Eq.~\ref{LDP5}, one obtains
\[
\Prob \Big\{x_{max} < 1+\tau -\frac{u}{\sqrt{N}}\Big\} \approx \exp \Big({-a N^{2-\frac{p}{2}} u^{p} - b N^{1-\frac{q}{2}} u^{q}}\Big)
\]
It is apparent that this yields a nontrivial finite shape only if $p=4$ and $q=2$.}
 that in the limit $x\to 1+\tau-0$
\begin{equation}\label{LDP5}
K_{\tau}(x) \sim a_K\,  (1+\tau-x)^{p_K}, \quad T_{\tau} (x) \sim b_T\,   (1+\tau-x)^{q_T}\, ,
\end{equation}
where $a_K,b_T,p_K,q_T>0$.

Now, since the integral in Eq.~\ref{NstassB} is dominated by the immediate neighborhood of its upper limit of integration at $x=1+\tau$, then to leading order in $N$
\[
\frac{1}{N}\ln I_{st}{(N)} = \frac{1}{N}\, \ln \int_{1+\tau-\varepsilon}^{1+\tau}  e^{N{\Sigma}_{st} (x)-N^2 K_{\tau} (x) -N T_{\tau} (x) } dx \, ,
\]
where $\varepsilon$ can be chosen arbitrary small, and (by making us of Eq.~\ref{LDP5} and expanding $ {\Sigma}_{st} (x)$ in powers of $t=1+\tau-x$)
\begin{equation}\label{60}
\frac{1}{N}\ln  I_{st}{(N)} =
{\Sigma}_{st} (1+\tau) + \frac{1}{N}\, \ln \int_{0}^{\varepsilon} e^{-a_K N^2 t^{p_K} +c N t^q} dt \, ,
\end{equation}
where $q>0$. Regardless of the sign of $c$ the integral on the right-hand side has at most sub-exponential growth with $N$.
\begin{proposition} \label{Prop61} Let
\[
I(N; a,b; p,q) = \int_0^{\varepsilon} e^{-a N^2 t^{p} +c N t^q} dt\, \quad (a, p,q>0).
\]
Then
\begin{equation}\label{I_N}
\lim_{N\to\infty} \frac{1}{N}\ln  I(N; a,b; p,q)  =0\,  .
\end{equation}
\end{proposition}

\emph{Proof}  If $c\le 0$ then
\[
I(N;a,b; p,q) \le  \int_0^{\varepsilon} e^{-a N^2 t^{p}} dt\, <  \frac{1}{N^{2/p}} \int_0^{+\infty} e^{-a s^{p}} ds,
\]
and, obviously, Eq.~\ref{I_N} holds. Similarly, if $c>0$ and $p=q$ then for $N$ sufficiently large
\[
I(N;a,b; p,q) = \int_0^{\varepsilon} e^{-(a N^2 -cN) t^{p}} dt < \frac{1}{(a N^2 -cN)^{1/p}} \int_0^{+\infty} e^{- s^{p}} ds
\]
and Eq.~\ref{I_N} holds. It remains to consider $c>0$ and $p\not=q$. By changing the variable of integration from $t$ to $s=t/N^{1/(p-q)}$,
\[
I(N;a,b; p,q) = \frac{1}{N^{\frac{1}{p-q}}}\int_0^{\varepsilon N^{\frac{1}{p-q}}} e^{-N^{\frac{p-2q}{p-q}}\varphi (s)} ds\, , \quad \text{where } \varphi (s)= a s^p -c s^{q}\, .
\]
Note that the function $\varphi (s)$ vanishes on the interval $0\le s< \infty$ at two points only, $s=0$ and $s_1=(c/a)^{1/(p-q)}$. It is negative on the interval $0<s<s_1$ where it has one point of minimum at $s=s_m$ and positive for all $s>s_1$. Therefore,
\[
I(N;a,b; p,q) = \frac{1}{N^{\frac{1}{p-q}}}\Big( \int_0^{s_1}+ \int_{s_1}^{\varepsilon N^{\frac{1}{p-q}}} \Big)\,  e^{-N^{\frac{p-2q}{p-q}}\varphi (s)} ds < \frac{s_1e^{N^{\frac{p-2q}{p-q}}|\varphi (s_m)|} }{N^{\frac{1}{p-q}}} \, + \,  \frac{\varepsilon N^{\frac{1}{p-q}} - s_1}{N^{\frac{1}{p-q}}}\, ,
\]
and Eq.~\ref{I_N} holds. This proves that assertion of Proposition \ref{Prop61}.  \hfill $\Box$

\medskip

Now, it follows Eq.~\ref{60} and Proposition \ref{Prop61} that 
\begin{equation}\label{62}
\lim_{N\to\infty}\frac{1}{N}\ln   I_{st}{(N)} =
{\Sigma}_{st} (1+\tau)= \lim_{x\to1+\tau-0}\mathit{\Phi} (x; \nu_x) -\frac{(1+\tau-m)^2}{2\tau} -\ln m \, .
\end{equation}
This is in parallel with Eq.~\ref{NstassAfin},
\[
\lim_{N\to\infty}\frac{1}{N}\ln   I_{eq}{(N)}  =
{\Sigma}_{eq} (1+\tau)= \lim_{x\to1+\tau+0}\mathit{\Phi} (x; \mu_{eq}) -\frac{(1+\tau-m)^2}{2\tau} -\ln m \, .
\]
It is now apparent that under the very natural assumption of continuity of the logarithmic potential, 
\begin{equation}\label{cont}
\tcboxmath{
\text{Assumption: } \quad \lim_{x\to1+\tau-0}\mathit{\Phi} (x; \nu_x)=\lim_{x\to1+\tau+0}\ \mathit{\Phi} (x; \mu_{eq})\,
}
\end{equation}
 both integrals have the same rate of exponential growth. As $\mathit{\Phi} (1+\tau; \mu_{eq})=\frac{\tau}{2}
$, we then conclude that
\begin{equation}\label{mainfinding}
\tcboxmath{
\lim_{N\to\infty} \frac{1}{N}\ln \langle {\cal N}_{st}\rangle = \mathit{\Sigma}_{st}\,  \quad \mathit{\Sigma}_{st}=-\left[1-m+\ln{m}+\frac{(m-1)^2}{2\tau}\right] \quad\quad (0<m<1)
}
\end{equation}
as was claimed in our paper. It is straightforward to see that
\begin{equation}\label{mainfindingA}
\mathit{\Sigma}_{st}=\mathit{\Sigma}_{eq} - \frac{1+\tau}{2\tau}\, (m-1)^2 \quad\quad (0<m<1)\, .
\end{equation}
Hence, in the annealed approximation the probability for a randomly selected equilibrium to be stable is exponentially small
\[
p^{(a)}_{st}= \frac{\langle {\cal N}_{st}\rangle }{\langle {\cal N}_{eq}\rangle } \approx  e^{-N \frac{1+\tau}{2\tau}\, (m-1)^2}\quad \quad (0<m<1)\, ,
\]
i.e.,
\[
\lim_{N\to\infty} \frac{1}{N}\ln p^{(a)}_{st} = -\frac{1+\tau}{2\tau}\, (m-1)^2\, .
\]

\section{Statistics of $\alpha-$stable equilibria} \label{subsection_G}

In this Section we evaluate the average number of $\alpha$-stable equilibria $\langle{\cal N}_{st}^{(\alpha)}\rangle$ and their relative density $\nu^{(a)}_N(\alpha)$, see Eqs.~\ref{denind} and \ref{A:Nstmain_intro} in the limit $N\gg1$.

\medskip

Eq.~\ref{E:LDoutcome} suggests splitting the interval of integration in Eq.~\ref{A:Nstmain_intro}  in two domains $x>x(\alpha)$ and $x<x(\alpha)$, where $x(\alpha)$ is the solution of Eq.~\ref{x_alpha} for $x$. Note that
\[
|x(\alpha)|\le 1+\tau\, .
\]
Correspondingly,
\begin{equation}\label{Nstass}
\langle{\cal N}_{st}^{(\alpha)}\rangle= I_{eq}^{(\alpha)}{(N)} +I_{st}^{(\alpha)}{(N)}\, .
\end{equation}
Here
\begin{align}\label{NstassA_alpha}
I_{eq}^{(\alpha)}{(N)} &= \int_{x(\alpha)}^{\infty} e^{N{\Sigma}_{eq}(x) +o(N)}\, dx , & \\[1ex]
\label{NstassB_alpha}
I_{st}^{(\alpha)}{(N)} &= \int_{-\infty}^{x(\alpha)} e^{N{\Sigma}_{st}^{(\alpha)} (x)+o(N)} \Prob\{ x_{\alpha N+1} <x \}
\, dx, &\text{where} \quad{\Sigma}_{st}^{(\alpha)} (x)=\mathit{\Phi }(x;\nu_x^{(\alpha)}) - \frac{(x-m)^2}{2\tau} -\ln m \, .
\end{align}
Both integrals can be evaluated in the limit $N\gg 1$ by the Laplace method.

\medskip

\begin{figure}[t!]
\centering
\includegraphics[width=.5\linewidth]{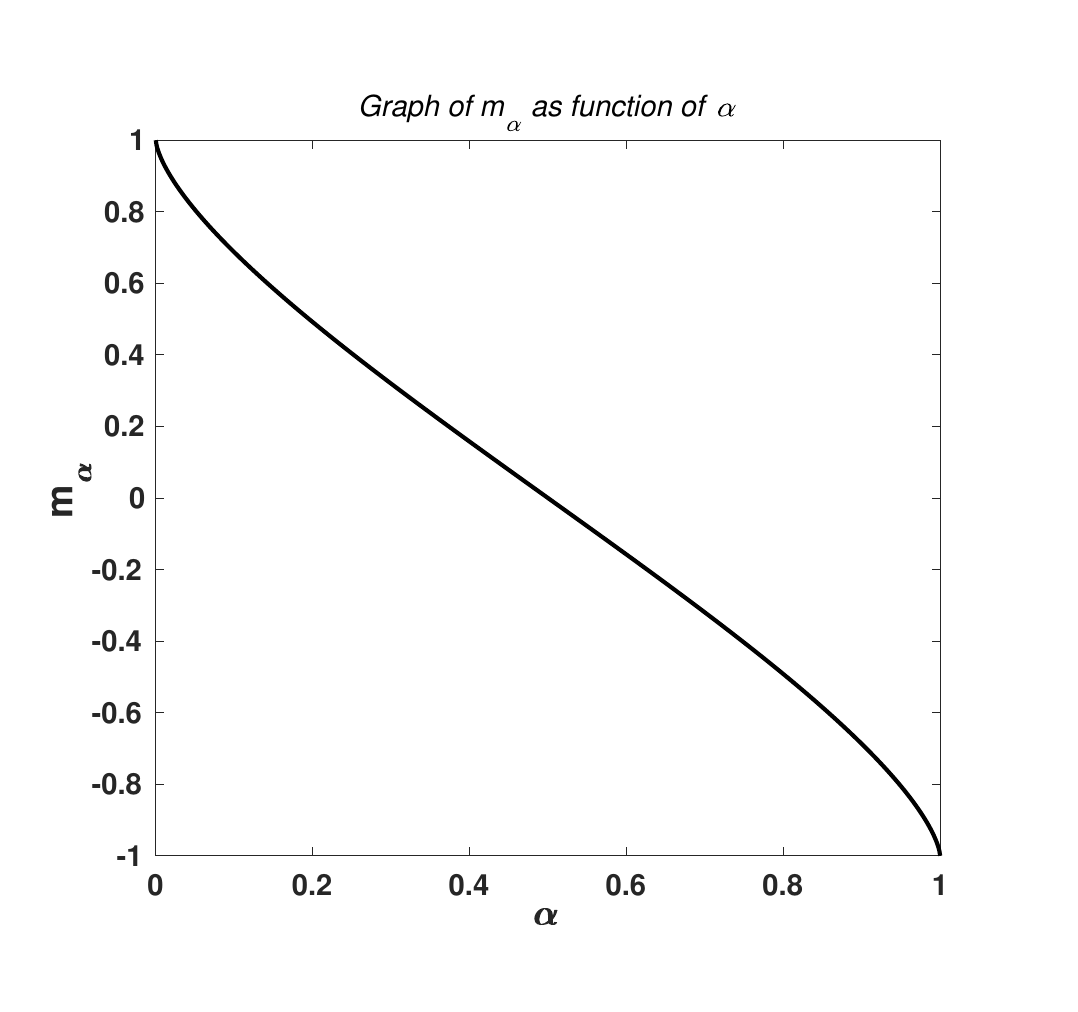}
\vspace*{-3ex}
\caption{The graph of $m_{\alpha}$ as function of $\alpha$. This is fairly linear function with a small curvature at the end points, taking values 1, 0 and -1 at $\alpha=0, 0.5, 1$ correspondingly. }
\label{fig:2}
\end{figure}

Recall that the function ${\Sigma}_{eq}(x) $ has global maximum at $x_{*}=(1+\tau)m$, $|x_{*}|\le 1+\tau$, being monotone increasing on the interval $x< x_{*}$ and monotone decreasing on the interval  $x>x_{*}$. Since the integral in Eq.~\ref{NstassA_alpha} is dominated by immediate neighborhood of the point of maximal value of ${\Sigma}_{eq}(x) $ on the interval of integration, the rate of exponential growth of $I_{eq}^{(\alpha)}{(N)} $ will depend on the position of the lower boundary of integration $x(\alpha)=(1+\tau)m_{\alpha}$ relative to $x_{*}=(1+\tau)m$, see Fig.~\ref{fig:1}:
\begin{equation}\label{a1}
I_{eq}^{(\alpha)}{(N)}\approx
 \begin{cases}
 e^{N {\Sigma}_{eq}((1+\tau)m_{\alpha})},  & \text{if } 0<m<m_{\alpha} \, ,\\[1ex]
 \displaystyle{
e^{N {\Sigma}_{eq}((1+\tau)m) }
 }  \, ,
 &  \text{if } m_{\alpha}<m<1 \, ,
 \end{cases}
\end{equation}
where  $m_{\alpha}$ is the unique solution of the equation
\begin{equation}\label{m_alpha}
\tcboxmath{
\text{Equation for $m_{\alpha}$:  }\quad \alpha =   \frac{2}{\pi} \int_{m}^1 \sqrt{ 1 - t^2  } \, dt \quad \quad (0<\alpha<1)\,
}
\end{equation}
for $m$ in the interval $|m|\le 1$, see Fig.~\ref{fig:2}.

The integral in Eq.~\ref{NstassB_alpha} can be evaluated in the limit $N\gg 1$ by copying the evaluation of $I_{st}(N)$ in Section \ref{subsection_F} . Under the assumptions that
\begin{tcolorbox}
\begin{itemize}
\item[(a)] the next sub-leading term in the exponential in the large deviation principle for $\Prob \{ x_{\alpha N+1} <x \}$ is of order $N$,
\begin{equation*}
\Prob \{ x_{\alpha N+1}  <x \} = e^{-N^2K_{\tau}^{\alpha}(x)-NT_{\tau}^{\alpha}(x)+o(N)}\,   \quad \quad (x<1+\tau).
\end{equation*}
\item[(b)]  the log-potential is continuous
\begin{equation*}
\lim_{x\to x(\alpha)-0}\mathit{\Phi} (x; \nu^{(\alpha)}_x)=\lim_{x\to x(\alpha) +0}\ \mathit{\Phi} (x; \mu_{eq})\,
\end{equation*}
\end{itemize}
\end{tcolorbox}
it can be shown that
\begin{equation}\label{a2}
I_{st}^{(\alpha)}{(N)}  \approx e^{N {\mathit{\Sigma}}_{eq}((1+\tau)m_{\alpha})}\quad \quad (0<m<1)\, .
\end{equation}
It is easy to verify that
\begin{equation*}
 {\Sigma}_{eq}((1+\tau)m) = \mathit{\Sigma}_{eq} \quad \text{ and }\quad  {\Sigma}_{eq}((1+\tau)m_{\alpha})= \mathit{\Sigma}_{eq} -\frac{1+\tau}{2\tau}(m_{\alpha}-m)^2,
\end{equation*}
where $ \mathit{\Sigma}_{eq} $ is given by Eq.~\ref{Sigma_eq}. This together with Eq.~\ref{Nstass} and Eqs~\ref{a1}--\ref{a2} implies the desired result
\begin{equation}\label{mainfinding10}
\tcboxmath{
\lim_{N\to\infty} \frac{1}{N}\ln \langle{\cal N}_{st}^{(\alpha)}\rangle =\begin{cases}
 \mathit{\Sigma}^{(\alpha)}_{st} & \text{if } 0<m<m_{\alpha} \, ,\\[1ex]
\mathit{\Sigma}_{eq}  \, ,
 &  \text{if } m_{\alpha}<m<1 \, ,
 \end{cases}
  }
\end{equation}
where
\begin{equation}\label{Sigma_alpha}
\mathit{ \Sigma}^{(\alpha)}_{st} = \mathit{\Sigma}_{eq} -\frac{1+\tau}{2\tau}(m_{\alpha}-m)^2\, .
 \end{equation}

\medskip

Let us fix now $m\in (0,1)$ and $\tau\in [0,1)$ and consider the relative {\it density} of $\alpha$-stable equilibria, Eq.~\ref{denind}.
Since $m\in (0,1)$ we have $ \mathcal{N}_{st}= \mathcal{N}^{(0)}_{st} \ll  \mathcal{N}^{(1)}_{st}=\mathcal{N}_{eq}$, and in the limit $N\gg1$:
\begin{equation}\label{normcond}
\int_0^1\nu^{(a)}_N(\alpha)\,d\alpha=\frac{\mathcal{N}_{eq}-\mathcal{N}_{st}}{\mathcal{N}_{eq}} \sim 1\, .
\end{equation}
Our next goal is to obtain this density in the limit $N\to\infty$. To this end, we first note that that Eqs~\ref{mainfinding10}--\ref{Sigma_alpha} 
in fact imply that for {\it every} $\alpha\in(0,1)$
\begin{equation}\label{alphadir}
\frac{d}{d\alpha}\left\langle \mathcal{N}^{(\alpha)}_{st}\right\rangle\approx
e^{N\left[\mathit{\Sigma}_{eq}(m)-\frac{1+\tau}{2\tau }\left(m_{\alpha}-m\right)^2 \right]}\, .
\end{equation}
This relation can be established by adjusting the asymptotic analysis of $\mathcal{D}_N^{(\alpha)}(x)$ and $\langle {\cal N}_{st}^{(\alpha)}\rangle$ to the computation of $\Delta  \langle {\cal N}_{st}^{(\alpha)}\rangle$,
\[
\Delta  \langle {\cal N}_{st}^{(\alpha)}\rangle =  \langle {\cal N}_{st}^{(\alpha+\Delta \alpha)}\rangle - \langle {\cal N}_{st}^{(\alpha)}\rangle \, .
\]
Indeed, consider fixed $\alpha \in (0,1)$ and $\Delta \alpha \ll \alpha$. Then
\[
\Delta  \langle {\cal N}_{st}^{(\alpha)}\rangle = \int_{-\infty}^{\infty} \!  \Delta \mathcal{D}_N^{(\alpha)}(x)\,  e^{-N \big[ \frac{(x-m)^2}{2\tau}+\ln m\big]} \frac{dx}{\sqrt{2\pi\tau/N}} \quad \quad (0<\alpha <1)
\]
where
\[
\Delta \mathcal{D}_N^{(\alpha)}(x) =  \left\langle \Delta \mathit{\Theta} \,   |\det (X-xI)| \right\rangle_X\, , \quad \Delta \mathit{\Theta} := \mathit{\Theta} \big(\alpha +\Delta\alpha -\mu_N(\mathbb{H}_x)\big) -\mathit{ \Theta} \big(\alpha -\mu_N(\mathbb{H}_x)\big)\, .
\]
Now note that $\mu_N(\mathbb{H}_x)$ counts the proportion of the eigenvalues of $X$ located to the right of the vertical line $\Re z =x$ so that $\Delta \Theta$ is the indicator function of the event that $\mu_N(\mathbb{H}_x)\in (\alpha, \alpha +\Delta\alpha) $ and
\[
\left\langle \Delta \mathit{\Theta} \right\rangle_X = \Prob\{\mu (\mathbb{H}_x) \in ( \alpha,\alpha +\Delta\alpha )   \} \, .
\]
Exploiting the large deviation principle for the empirical eigenvalue counting measure $ \mu_N$, much in the same way as in Section \ref{subsection_E} one obtains the factorization  $\Delta \mathcal{D}_N^{(\alpha)}(x) $ into the product of $\left\langle \Delta \mathit{\Theta} \right\rangle_X $ and $e^{N\mathit{\Phi} (x; \mu_x^{\alpha})}$ where now $ \mu_x^{\alpha}$ is the minimizer of the large deviation rate functional $J_{\tau}[\mu]$ on the set of symmetric probability measures on $\mathbb{C}$ satisfying the condition $ \mu (\mathbb{H}_x) \in [ \alpha,\alpha +\Delta\alpha]$ with $e^{N\mathit{\Phi} (x; \mu_x^{\alpha})}$ being its log-potential. It is then follows that
\[
\Delta  \langle {\cal N}_{st}^{(\alpha)}\rangle \approx  \int_{-\infty}^{\infty} \! e^{-N^2 J_{\tau}[\mu_x^{\alpha}] + N\big[\mathit{\Phi} (x; \mu_x^{\alpha})  - \frac{(x-m)^2}{2\tau}-\ln m \big]} \, dx
\]
This integral is dominated by the immediate neighborhood of $x=x(\alpha)$ where $J_{\tau}[\mu_x^{\alpha}] $ as a function of $x$ is vanishing, $J_{\tau}[\mu_{x(\alpha)}^{\alpha}]=0$. And under assumptions similar to (i) and (ii) in Section \ref{subsection_E} (namely, (i) the next-to-leading term in the large deviation principle is of order $N$ and (ii) the log-potential $\mathit{\Phi} (x; \mu_x^{\alpha})$ is continuous function of $x$ at $x=x(\alpha)$ ) one obtains Eq.~\ref{alphadir}.

\medskip

Eq.~\ref{alphadir} immediately implies that
\begin{equation}\label{alphadir1}
\tcboxmath{
\nu^{(a)}_N(\alpha)\approx e^{-N\frac{1+\tau}{2\tau}(m_{\alpha}-m)^2}\, .
}
\end{equation}
Even though the large deviations technique does not allow to control pre-exponential factors, on this occasion, the pre-exponential factor can be restored  from the normalization condition of Eq.~\ref{normcond}. To this end, let us write
\[
\nu^{(a)}_N(\alpha) =c_N(m) e^{-N\frac{1+\tau}{2\tau}(m_{\alpha}-m)^2} \quad \quad (0 \le \alpha \le 1).
\]
Substituting this into Eq.~\ref{normcond} and changing the variable of integration there from $\alpha$ to $u=m_{\alpha}$, see Fig.~\ref{fig:2}, one obtains the equation
\[
\frac{2}{\pi}\int_{-1}^1  c_N(m) \sqrt{1-u^2}\,  e^{-N\frac{1+\tau}{2\tau}(u-m)^2}\, du =1
\]
which must hold in the limit $N\gg 1$. Evaluating this integral by the Laplace method we determine the pre-exponential factor
\[
c_N(m)=\sqrt{\frac{N\pi(1+\tau) }{8\tau(1-m^2)}}\, .
\]
Thus, this procedure yields the density of the  number of unstable directions at typical equilibria of our model in the limit $N\gg 1$:
\begin{equation}\label{denalpha}
\tcboxmath{
\nu^{(a)}_N(\alpha)=\sqrt{\frac{N\pi (1+\tau)}{8\tau(1-m^2)}}\, e^{- \frac{1}{2}\frac{N(1+\tau)}{\tau}\left(m_{\alpha}-m\right)^2}\quad \quad (0 \le \alpha \le 1)\, .
}
\end{equation}

We see therefore that for a fixed value of the parameter $m\in(0,1)$, only indices $\alpha$ in a small interval of width of order $\sqrt{\tau/N}$ around the value $\alpha=\alpha_m$,
\begin{equation}\label{alpha_m}
\alpha_m = \frac{2}{\pi} \int_{m}^1\sqrt{1-t^2}\, dt,
\end{equation}
have finite densities.
In other words, the value of parameter $m$ (but not of $\tau$) dictates the most probable value of the instability index of a typical  equilibrium. The dependence of the most probable value of the instability index on $m$ is continuous. In particular, for values of $m$ in the topologically nontrivial phase close to the instability  threshold at $m=1$,
\begin{equation}\label{alpha_m1}
\alpha_m \sim  
\frac{4\sqrt{2} }{3\pi} \, \varepsilon^{3/2} \quad \quad  (\varepsilon= 1-m, \,  \varepsilon \ll 1) .
\end{equation}
This can be easily seen from Eq.~\ref{alpha_m}. In the context of the May-Wigner instability transition, once the system complexity exceeded the critical value and the system transitioned into the topologically nontrivial phase with exponentially many equilibria which are typically all unstable, the instability index, i.e. the proportion of unstable directions, of a typical equilibrium remains low but increasing as the complexity of the system increases. One interesting question is about how many unstable directions would a typical equilibrium have for parameter value $m$ in the transition region from stability to instability. The width of this transition region is $N^{-1/2}$ and our technique of large deviations does not give access to $\nu^{(a)}_N(\alpha)$ in this region in its entirety. However, one would reasonably expect that the density of the unstable directions in the left tail of the transition matches the expression in Eq.~\ref{denalpha}. Correspondingly, we set
\begin{equation}\label{a8}
m=1-\frac{\delta}{\sqrt{N}}, \quad \text{where } 1\ll \delta \ll \sqrt{N}.
\end{equation}
It is apparent that for such values of $m$,  the density $\nu^{(a)}_N(\alpha)$ is not vanishing in the limit $N\gg 1$ only for small values of $\alpha$. By inverting the relation between $\alpha$ and $m_{\alpha}$ in Eq.~\ref{alpha_m1} one obtains
\begin{equation}\label{a9}
m_{\alpha}=1- \left(\frac{3\pi}{4\sqrt{2}}\, \alpha \right)^{2/3} \quad\quad  (\alpha \ll 1).
\end{equation}
It is now apparent from Eqs~\ref{denalpha}, \ref{a8} and \ref{a9} that in the left tail of the transition region the instability index $\alpha$ scales as
$
\alpha={\gamma}/{N^{3/4}}
$,
and, hence,
\[
m_{\frac{\gamma}{N^{3/4}}}= 1- \left(\frac{3\pi}{4\sqrt{2}}\, \gamma \right)^{2/3}\!\! \frac{1}{\sqrt{N}}\, .
\]
On substituting this and Eq.~\ref{a8} into Eq.~\ref{denalpha} one obtains the desired density of the unstable directions at typical equilibria in the left tail of the transition region:
\begin{equation}\label{a10}
\tcboxmath{
\left.\frac{1}{N^{3/4}}\nu^{(a)}_N\Big(\frac{\gamma}{N^{3/4}}\Big)\right|_{m=1- \frac{\delta}{\sqrt{N}}}= \sqrt{\frac{\pi(1+\tau)}{16\, \tau \delta}}\, e^{
-\frac{1+\tau}{2\tau}
\big[
\delta-\frac{1}{2}
\big( \frac{3\pi}{2}\gamma
\big)^{2/3}
\big]^2
 }
 \quad (1\ll \delta \ll \sqrt{N})
 }
\end{equation}
It is instructive to verify the density of the unstable directions given by Eq.~\ref{a10} integrates to 1 in the limit $\delta \gg 1$. We have
\begin{eqnarray*}
\int_0^1\nu^{(a)}_N(\alpha)\, d\alpha &=& \int_0^{N^{3/4}} \frac{1}{N^{3/4}}\nu^{(a)}_N\Big(\frac{\gamma}{N^{3/4}}\Big)\, d\gamma \\
&\sim &  \sqrt{\frac{\pi(1+\tau)}{16\tau\delta}}  \int_0^{+\infty} \!
e^{
-\frac{1+\tau}{2\tau}
\big[
\delta-\frac{1}{2}
\big( \frac{3\pi}{2}\gamma
\big)^{2/3}
\big]^2
 }  \, d\gamma = \sqrt{\frac{1+\tau}{2\pi\tau}}\int_{-\delta}^{+\infty} \sqrt{1+\frac{x}{\delta}}\, e^{-\frac{1}{2} \frac{1+\tau}{\tau}x^2}\, dx\, .
\end{eqnarray*}
where we have changed the variable of integration from $\gamma$ to $x=\left(\frac{3\pi}{4\sqrt{2}}\, \gamma \right)^{2/3}-\delta$. It is evident that to leading order in the limit $\delta \gg 1$ the integral over $x$ above is 1, as is expected.

\bibliography{BFK}